\documentclass[letterpaper,12pt]{article}
\usepackage[pass,paperwidth=8.5in,paperheight=11in,top=1in,left=1in,bottom=1in,right=1in]{geometry} 
\usepackage[OT1]{fontenc}
\usepackage{amsthm}
\usepackage{amsmath}
\usepackage{amssymb}
\usepackage[numbers]{natbib}
\usepackage[colorlinks,citecolor=blue,urlcolor=blue]{hyperref}
\usepackage{epstopdf}
\usepackage{dsfont}
\usepackage{subfig}
\usepackage[dvips]{graphicx}
\usepackage{graphics}
\usepackage{authblk}
\usepackage{caption}
\usepackage{comment} 
\usepackage{appendix} 
\usepackage{setspace}
\usepackage{float}
\usepackage{lscape}
\restylefloat{table}
\restylefloat{figure}
\bibpunct{(}{)}{;}{a}{}{;}
\numberwithin{equation}{section}

\newtheorem{prop}{Proposition}[section]

\newcommand{\defeq}{\mathrel{\mathop:}=}

\title{Reduced-Rank Covariance Estimation \\ in Vector Autoregressive Modeling}
\date{}
\author[1]{Richard A. Davis\thanks{rdavis@stat.columbia.edu.}}
\author[1]{Pengfei Zang\thanks{pengfei.columbia@gmail.com. To whom correspondence should be addressed.}}
\author[1]{Tian Zheng\thanks{tzheng@stat.columbia.edu.}}
\affil[1]{Department of Statistics, Columbia University}

\begin{document} 
\maketitle
\begin{singlespace}
\begin{abstract} 
We consider reduced-rank modeling of the white noise covariance matrix in a large dimensional vector autoregressive (VAR) model. We first propose the reduced-rank covariance estimator under the setting where independent observations are available. We derive the reduced-rank estimator based on a latent variable model for the vector observation and give the analytical form of its maximum likelihood estimate.
Simulation results show that the reduced-rank covariance estimator outperforms two competing covariance estimators for estimating large dimensional covariance matrices from independent observations. Then we describe how to integrate the proposed reduced-rank estimator into the fitting of large dimensional VAR models, where we consider two scenarios that require different model fitting procedures. In the VAR modeling context, our reduced-rank covariance estimator not only provides interpretable descriptions of the dependence structure of VAR processes but also leads to improvement in model-fitting and forecasting over unrestricted covariance estimators. Two real data examples are presented to illustrate these fitting procedures.
~\\{\bf Keywords}: Covariance Estimation; Vector Autoregressive (VAR) Models; Matrix Decomposition.
\end{abstract}
\end{singlespace}

\section{Introduction}\label{section_introduction_rr}
Suppose $\{Y_{t}\} = \{(Y_{t,1},Y_{t,2},\ldots,Y_{t,K})^{'}\}$ is a $K$-dimensional stationary time series that follows the vector autoregressive model of order $p$ (VAR($p$))
\begin{eqnarray}
Y_{t}-\mu = \displaystyle\sum_{k=1}^{p}A_{k}(Y_{t-k}-\mu) + Z_{t},~t=1,\ldots, T, \label{VAR_equation}
\end{eqnarray}
where $\mu$ is a real-valued $K$-dimensional vector; $A_{1},\ldots,A_{p}$ are real-valued $K\times K$ matrices of autoregressive (AR) coefficients; and $\{Z_{t}\}$ is a sequence of iid $K\times 1$ noise with mean $\mathbf{0}$ and covariance matrix $\Sigma_{Z}$. We further assume that the process $\{Y_{t}\}$ is \textit{causal}, i.e., $\det(I_{K} - \displaystyle\sum_{k=1}^{p}A_{k}z^{k})\ne 0$, for $z \in \mathbb{C}, |z| < 1$, e.g., see \citet{Helmut1991}. The VAR model \eqref{VAR_equation} has been applied for modeling the joint evolution of multivariate series in many fields, such as political science \citet{Freeman1989}, macroeconomics \citet{Sims1980}, biological science \citet{Holter2001} and finance \citet{Eun1989}.

One indispensable aspect of fitting the VAR model \eqref{VAR_equation} is the estimation of the noise covariance matrix $\Sigma_{Z}$: an estimate of the noise covariance matrix $\Sigma_{Z}$ is needed for exploring the dependence structure of the VAR process \citet{Demiralp2003, Moneta2004} while an estimate of the inverse of the noise covariance matrix $\Sigma_{Z}^{-1}$ is required in constructing confidence intervals for AR coefficients or for computing the mean squared error of VAR forecasting \citet{Helmut1991}. A natural estimator for $\Sigma_{Z}$ in a VAR model is the sample covariance matrix of the residuals from fitting an autoregression \citet{Helmut1991}. To this end, the residuals are viewed as independent samples, conditioned on the AR coefficient estimates, from an underlying distribution with covariance matrix $\Sigma_{Z}$. Therefore estimating the noise covariance matrix in a VAR model can be cast as a covariance estimation problem where independent observations are available. 

Covariance estimation from independent observations is a fundamental problem in many areas, such as portfolio selection \citet{Ledoit2004}, functional genomics \citet{Schafer2005}, fMRI study \citet{Daniels2001} and graphical models \citet{Lauritzen1989}. Estimating a $K\times K$ covariance matrix posits many challenges for large $K$ since the number of parameters to be estimated $\frac{K(K+1)}{2}$ grows quadratically in the dimension $K$. The sample covariance matrix of the observations serves as a natural estimator when the dimension $K$ is much smaller than the sample size. But it is also well-known that the sample covariance matrix can be severely ill-conditioned in small- to medium- samples. As a result, various methods have been proposed to estimate large dimensional covariance matrices. The three most common approaches are {\em shrinkage}, where the covariance estimator is obtained by shrinking the sample covariance matrix towards a pre-specified covariance structure \citet{Ledoit2004, Schafer2005}; {\em regularization}, where the covariance estimator is derived based on regularization methods, such as banding \citet{Bickel2008}, thresholding \citet{Karoui2008} and penalized estimation \citet{Huang2006}; and {\em structural}, where structural constraints, such as factor structures \citet{Tipping1999} or autoregressive structures \citet{Daniels2001}, are imposed to reduce the effective dimension of the covariance estimator.

In this paper, we propose a reduced-rank estimator for the noise covariance matrix in a large dimensional VAR model. In Section \ref{section_RRintro_rr} we first derive the reduced-rank estimator under the setting when observations are independent. The reduced-rank estimator is based on a latent variable model for the data and its effective dimension can be much lower than the dimension of the population covariance matrix. So the reduced-rank estimator can be viewed as a structural covariance estimator. The reduced-rank estimator is attractive since it is not only well-conditioned, but also provides an interpretable description of the covariance structure. Simulation results show that the reduced-rank covariance estimator outperforms two competing shrinkage estimators for estimating large dimensional covariance matrices. In Section \ref{section_RRappToVAR_rr}, we proceed to the context of VAR modeling. We describe how to integrate the proposed reduced-rank estimator into the fitting of large dimensional VAR models, for which we consider two scenarios that require different model fitting procedures. The first scenario is that there are no constraints on the AR coefficients, for which the VAR model can be fitted using a 2-step method; while in the second scenario there exist constraints on the AR coefficients, where the VAR model needs to be fitted by an iterative procedure. In Section \ref{section_realExample_rr}, the reduced-rank covariance estimator is applied to the VAR modeling of  two real data examples. The first example is concerned with stock returns from S\&P 500 and the second example is a time series of temperatures in southeast China.

\section{Reduced-rank covariance estimation}\label{section_RRintro_rr}
We first derive the reduced-rank covariance estimator based on independent observations. Then we proceed to VAR modeling and describe how to integrate the reduced-rank estimator into the fitting of large dimensional VAR models.

\subsection{For independent observations}\label{section_RR_independent_rr}
We assume that $Z_{1},\ldots,Z_{T}$ are $T$ independent replicates from a $K$-dimensional Gaussian distribution with covariance matrix $\Sigma_{Z}$ \footnote{Here we make the assumption of Gaussianity. If $Z_{t}$ is non-Gaussian, our proposed reduced-rank covariance estimation method can still be applied, where the Gaussian likelihood is interpreted as a quasi-likelihood.}. Without loss of generality, we assume that $\{Z_{t}\}$ has mean zero. The problem of interest is to estimate $\Sigma_{Z}$, which can be large dimensional. To derive our covariance estimator, we further assume that each vector observation $Z_{t}$ follows the latent variable model
\begin{equation}
Z_{t} = U\delta_{t} + \varepsilon_{t},\mbox{ for } t=1,\ldots,T, \label{latentVarModel}
\end{equation}
where the latent variables $\delta_{t}~(t=1,\ldots,T)$ are independent replicates from a $d$-dimensional ($1\le d\le K-1$) Gaussian with mean $\bf{0}$ and a diagonal covariance matrix $\Lambda\defeq\mathrm{diag}\{\lambda_1,\ldots,\lambda_d\}$ ($\lambda_{1} > \lambda_{2} > \ldots > \lambda_{d}>0$); $U$ is a $K\times d$ column-orthonormal matrix, i.e., $U^{'}U = I_{d}$; and the errors $\varepsilon_{t}~(t=1,\ldots,T)$ are independent replicates from a $K$-dimensional Gaussian with mean $\mathbf{0}$ and {\em isotropic} covariance matrix $\mathrm{cov}(\varepsilon_{t})=\sigma^{2}I_{K}$. As shown in Section \ref{section_RRcompFA_rr}, this isotropy assumption of the covariance matrix $\mathrm{cov}(\varepsilon_{t})$ is important in ensuring the identifiability of the latent variance model \eqref{latentVarModel} under Gaussianity.
 
Under the latent variable model \eqref{latentVarModel}, the covariance matrix $\Sigma_{Z}$ is seen to be
\begin{equation}
\Sigma_{Z} = U\Lambda U^{'} + \sigma^2I_{K}. \label{reduceRankCovMatr}
\end{equation}
The first component $U\Lambda U^{'}$ in the decomposition \eqref{reduceRankCovMatr} has reduced-rank $d$ ($d<K$) and contains the core information about the dependence structure between the $K$ dimensions of $Z_{t}$. The second component $\sigma^2I_{K}$ has a sparse structure and accounts for unexplained variability in individual dimensions. The decomposition \eqref{reduceRankCovMatr} approximates the $K$-dimensional dependence structure encoded by $\Sigma_{Z}$ with a rank-$d$ matrix $U\Lambda U^{'}$. Such an approximation is useful for separating important dependence patterns from large dimensional noisy observations.

\subsubsection{Connection and distinction with factor models}\label{section_RRcompFA_rr} 
The motivation of the latent variable model \eqref{latentVarModel} is that the $K$-dimensional vector $Z_{t}$ can be related to a $d$-dimensional vector $\delta_{t}$ of latent (unobserved) variables through a column-orthonormal matrix $U$. With $d<K$, the latent variable $\delta_{t}$ provides a more parsimonious description of the dependence structure of $Z_{t}$. This motivation is similar to that of factor models, see e.g., \citet{Anderson2003}. In the factor model setup, the relation \eqref{latentVarModel} is also used to link the observation with the latent variable and the matrix $U$ is called the {\em factor loading}; but it is usually assumed that
the latent variable $\delta_{t}$ has an isotropic covariance matrix while the error $\varepsilon_{t}$ has a non-isotropic covariance matrix. It is known that factor models have identifiability issues. Specifically, for any $d\times d$ orthogonal matrix $C$, the pairs $(U, \delta_{t})$ and $(UC^{'}, C\delta_{t})$ will lead to two equivalent factor models. In contrast, identifiability is not an issue in our latent variable model \eqref{latentVarModel}. This is because in the latent variable model we make different assumptions on the covariance structures of the latent variable $\delta_{t}$ and the error $\varepsilon_{t}$, as summarized in Table \ref{latentVarModelVSfactorModel}. In the latent variable model, the covariance matrix of the vector $C\delta_{t}$ is $\mathrm{cov}(C\delta_{t}) = C\mathrm{diag}\{\lambda_1,\ldots,\lambda_d\}C^{'}$, which in general is not equal to the original covariance matrix $\mathrm{cov}(\delta_{t})=\mathrm{diag}\{\lambda_1,\ldots,\lambda_d\}$. So the two latent variable models corresponding to the pairs $(U, \delta_{t})$ and $(UC^{'}, C\delta_{t})$ are not equivalent; in other words, the assumption of a non-isotropic covariance matrix for the latent vector $\delta_{t}$ leads to the identifiability of the latent variable model \eqref{latentVarModel}. As a result, interpretation of the matrix parameter $U$ becomes meaningful.

\begin{table}[hbt]
\caption{Comparison of assumptions between the latent variable model and the factor model.}
\label{latentVarModelVSfactorModel}
\center 
\begin{tabular}{l|c|c} 
Model                                                           & $\mathrm{cov}(\delta_{t})$                                      & $\mathrm{cov}(\varepsilon_{t})$ \\
\hline
latent variable model \eqref{latentVarModel}    & $\mathrm{diag}\{\lambda_1,\ldots,\lambda_d\}$ & $\sigma^2I_{K}$ \\
\hline
factor model                                                  & $\sigma^2I_{d}$                                                      & $\mathrm{diag}\{\lambda_1,\ldots,\lambda_K\}$ \\
\hline
\end{tabular}
\end{table}
 
\subsubsection{Maximum likelihood estimation}\label{section_RRmle_rr}
We derive the maximum likelihood estimator of the reduced-rank covariance matrix $\Sigma_{Z}=U\Lambda U^{'} + \sigma^2I_{K}$ \eqref{reduceRankCovMatr}. Based on observations $Z_{1},\ldots,Z_{T}$, $-\frac{2}{T}$log-likelihood, ignoring an additive constant, is given by
\begin{eqnarray}
-\frac{2}{T}\log L(U,\Lambda, \sigma^2) = \log|\Sigma_{Z}| + \mathrm{tr}(\Sigma_{Z}^{-1}S), \label{logLikelihood}
\end{eqnarray}
where $S\defeq\frac{1}{T}\displaystyle\sum_{t=1}^{T}Z_{t}Z_{t}^{'}$. The following proposition shows that there exists an analytical form for the maximum likelihood estimator of the reduced-rank covariance matrix $\Sigma_{Z}$. 
\begin{prop}\label{reduceRankDecomEstProp}
Let $c_{1}\ge c_{2}\ldots\ge c_{K}\ge0$ be the eigenvalues of the sample covariance matrix $S$ and assume that the reduced-rank $d$ is known. The maximum likelihood estimator of the reduced-rank covariance matrix $\Sigma_{Z}$ is given by
\begin{eqnarray}
\hat{\Sigma}_{Z} = \hat{U}\hat{\Lambda}\hat{U}^{'} + \hat{\sigma}^2 I_{K}, \label{reduceRankDecomEst}
\end{eqnarray}
where
\begin{align}
\hat{U} &= (\hat{U}_{1},\ldots,\hat{U}_{d}),\mbox{ and }\hat{U}_{i}\mbox{ is the eigenvector of } S \mbox{ corresponding to } c_{i}; \label{Uest} \\
\hat{\sigma}^2 &= \frac{1}{K-d}\displaystyle\sum_{i=d+1}^{K}c_i; and \label{mlesigma2}\\
\hat{\Lambda} &= \mathrm{diag}\{\hat{\lambda}_1,\ldots,\hat{\lambda}_d\}, \mbox{ with } \hat{\lambda}_i = c_i - \hat{\sigma}^2, ~i=1,\ldots,d.\label{mlelambda2}
\end{align}
\end{prop} 
\noindent We defer the proof to the Appendix \ref{section_appendix_proof_rr}.
 
\subsubsection{Properties of the reduced-rank covariance estimator}
From \eqref{Uest} we can see that there exist links between the latent variable model \eqref{latentVarModel} and {\em principal component analysis} (PCA), which is perhaps the most widely used statistical tool for dimension reduction. The common setup of PCA is based on a series of mutually-orthogonal projections of vector observations that maximize the retained variance, where the directions of these projections are called {\em principal axes}, see e.g., \citet{Jolliffe2002}. This setup is {\em not} based on a probabilistic model but comes from a projection perspective.
In contrast, the latent variable model \eqref{latentVarModel} provides a model-based formulation of PCA, in which the principal axes coincide with the columns of the maximum likelihood estimator $\hat{U}$ as given by \eqref{Uest}. In the literature, such a probabilistic formulation of PCA was first investigated by \citet{Lawley1953} within the context of factor analysis and was then studied by \citet{Tipping1999} under {\em probabilistic principal component analysis} (PPCA). A discussion on the advantages of this probabilistic formulation of PCA over the traditional projection-based setup is given in \citet{Tipping1999}.

We also investigate the conditioning property of the reduced-rank estimator $\hat{\Sigma}_{Z}$ \eqref{reduceRankDecomEst}. It can be shown that the eigenvalues, denoted by $b_{i}$ ($i=1,\ldots,K$), of the reduced-rank estimator $\hat{\Sigma}_{Z}$ are
\begin{eqnarray*}
 b_{i}  = \left\{ 
\begin{array}{lll}
        \hat{\lambda}_i + \hat{\sigma}^2 & = c_{i},                                                             & \mbox{for $i=1,\ldots,d$},\\
        ~~~~\hat{\sigma}^2                                & = \frac{1}{K-d}\sum_{i=d+1}^{K}c_i, & \mbox{for $i=d+1,\ldots,K$},
\end{array}\right. 
\end{eqnarray*}
which means that the reduced-rank estimator $\hat{\Sigma}_{Z}$ retains the $d$ largest eigenvalues but shrinks the remaining $(K-d)$ eigenvalues of $S$ towards their average. Therefore, the {\em condition number}, i.e., the ratio between the largest and smallest eigenvalues of the covariance estimator, of the reduced-rank estimator is smaller and often much smaller than that of the sample covariance matrix. In other words, the reduced-rank estimator can be better conditioned than the sample covariance matrix. In addition, as long as the reduced-rank $d$ is smaller than the sample size $T$, the reduced-rank estimator will be invertible even if the dimension $K$ exceeds the sample size $T$. 

Next we discuss how to control the complexity of a reduced-rank covariance estimator through the choice of its reduced-rank $d$. From \eqref{reduceRankDecomEst} we can see that there exist two extremes for $\hat{\Sigma}_{Z}$ as the reduced-rank $d$ varies: when $d=K-1$, i.e., there is no dimension reduction, $\hat{\Sigma}_{Z} = S$ becomes the full covariance model; and when $d=0$, i.e., there is no structured component $\hat{U}\hat{\Lambda}\hat{U}^{'}$, $\hat{\Sigma}_{Z} = \bar{c}I_{K}$ becomes the isotropic covariance model. In other words, the reduced-rank covariance estimator is obtained by balancing between the unbiased but highly variable sample covariance matrix and the biased but well-conditioned isotropic covariance matrix, where the balance is controlled by the reduced-rank $d$. In practice, the reduced-rank $d$ is unknown and needs to be estimated from data. Here we use the {\em Bayesian information criterion} (BIC), e.g., see \citet{Schwarz1978}, to determine the reduced-rank $d$. The BIC is computed as
\begin{eqnarray}\label{rr_BIC}
\mathrm{BIC}(d) &=& -2\log L(\hat{U},\hat{\Lambda},\hat{\sigma}^2) + \log(T)\times (Kd - \frac{d(d-1)}{2} + 1),
\end{eqnarray} 
where $L(\hat{U},\hat{\Lambda},\hat{\sigma}^2)$ is the maximized likelihood and $Kd - d(d-1)/2 + 1$ is the number of free parameters in the reduced-rank covariance estimator. We select the reduced-rank $d$ from $\{1,2,\ldots,K-1\}$ according to a minimum BIC. \citet{Tipping1999} give similar results on controlling the complexity of PPCA.

Finally we describe a diagnostic tool for the reduced-rank covariance model. The latent variable $\delta_{t}$ in \eqref{latentVarModel} can be estimated by
\begin{eqnarray}
\hat{\delta}_{t} = \hat{U}^{'}Z_{t},\mbox{ for }t=1,\ldots,T,\label{reduceRankDecomEstDelta}
\end{eqnarray}
where $\hat{U}$ is given by \eqref{Uest}. According to model assumptions, $\hat{\delta}_{1},\ldots,\hat{\delta}_{T}$ should behave like independent replicates from a $d$-dimensional Gaussian with a {\em diagonal} covariance matrix. So correlation functions of the estimated latent variable $\hat{\delta}_{t}$ \eqref{reduceRankDecomEstDelta} can be used for model diagnostics.

\subsection{For VAR series}\label{section_RRappToVAR_rr} 
In this section, we proceed from the setting of independent observations to VAR processes and apply the reduced-rank covariance estimator to the noise covariance matrix $\Sigma_{Z}$ in a VAR model \eqref{VAR_equation}.

As described in Section \ref{section_introduction_rr}, the reduced-rank estimator for $\Sigma_{Z}$ in a VAR model is computed based on the residuals from fitted autoregression. Therefore, in order to apply the reduced-rank covariance estimator, we need to estimate the AR coefficient matrices $A_{1},\ldots,A_{p}$ in \eqref{VAR_equation} as well, for which we consider two scenarios. The first scenario is that there are no constraints on the AR coefficient matrices $A_{1},\ldots,A_{p}$; while the second scenario is that there exist constraints on the AR coefficients. The second scenario occurs, for example, when some of the AR coefficients are constrained be to zero. Such zero constraints on AR coefficients arise when we model {\em Granger causality} of $\{Y_{t}\}$, see e.g., \citet{Granger1969}, or when we fit {\em sparse vector autoregressive} models to $\{Y_{t}\}$, see e.g., \citet{Davis2012}. Here we use zero constraints on AR coefficients as the example of the second scenario. Zero constraints on the AR coefficient matrices $A_{1},\ldots,A_{p}$ can be expressed as
\begin{equation}
\alpha\defeq\mathrm{vec}(A_{1},\ldots,A_{p}) = R\gamma, \label{sVARconstraint}
\end{equation}
where $\alpha\defeq\mathrm{vec}(A_{1},\ldots,A_{p})$ is the $K^2p$-dimensional vector obtained by stacking the columns of the AR coefficient matrices $A_{1},\ldots,A_{p}$; $R$ is a $K^2p\times m$ matrix of known constants with rank $m$; and $\gamma$ is a $m$-dimensional vector of unknown parameters. The matrix $R$ is referred to as the {\em constraint matrix} \citet{Davis2012} and it specifies which AR coefficients are zero by choosing one entry in each column to be $1$ and all the other entries in that column to be $0$. The rank $m$ of the constraint matrix $R$ is equal to the number of non-zero AR coefficients. Using results on constrained VAR estimation in \citet{Helmut1991} and on the reduced-rank covariance estimation in Section \ref{section_RR_independent_rr}, it can be shown that, under the constraint \eqref{sVARconstraint} and the reduced-rank covariance model \eqref{reduceRankCovMatr}, the maximum likelihood estimator of the AR coefficients $\alpha$ is given by
\begin{align}
\hat{\alpha} &= R[R^{'}(LL^{'} \otimes \hat{\Sigma}_{Z}^{-1})R]^{-1}R^{'}(L\otimes \hat{\Sigma}_{Z}^{-1})y, \label{sVAR_estAR}
\end{align}
where 
\begin{align*}
L_{t}      &\defeq (Y_{t}, Y_{t-1},\ldots, Y_{t-p+1})^{'},\\
L &\defeq (L_{0}, L_{1},\ldots, L_{T-1}),\\
y            &\defeq\mathrm{vec}(Y) = \mathrm{vec}(Y_{1}, Y_{2},\ldots, Y_{T}),
\end{align*}
and $\hat{\Sigma}_{Z}$ in \eqref{sVAR_estAR} is the reduced-rank maximum likelihood estimator for the noise covariance matrix $\Sigma_{Z}$ based on the residuals $\hat{Z}_{t}\defeq Y_{t}-\displaystyle\sum_{k=1}^{p}\hat{A}_{k}Y_{t-k}~ (t=p+1,\ldots,T)$ from the fitted autoregression.

~\\{\bf The model fitting procedure for the first scenario}.\\
When there are no constraints on the AR coefficients (scenario 1), we have $R=I_{K^2p}$ in \eqref{sVARconstraint} and \eqref{sVAR_estAR} becomes
\begin{eqnarray}
\hat{\alpha} &=& I_{K^2p}[I_{K^2p}^{'}(LL^{'} \otimes \hat{\Sigma}_{Z}^{-1})I_{K^2p}]^{-1}I_{K^2p}^{'}(L\otimes \hat{\Sigma}_{Z}^{-1})y \nonumber \\
&=& [(LL^{'})^{-1}\otimes \hat{\Sigma}_{Z}](L\otimes \hat{\Sigma}_{Z}^{-1})y \nonumber \\
&=& [(LL^{'})^{-1}L\otimes I_{K}]y. \label{fullVAR_estAR}
\end{eqnarray}
So for the first scenario, \eqref{fullVAR_estAR} shows that the estimation of the AR coefficients $\alpha$ does not involve the reduced-rank estimation of the noise covariance matrix $\Sigma_{Z}$. Therefore the reduced-rank covariance estimator can be applied to a VAR model using the following 2-step method.
\begin{itemize}
\item [Step 1.] Fit an unconstrained VAR model to $\{Y_{t}\}$ and obtain the AR coefficient estimates $\hat{\alpha}$ according to \eqref{fullVAR_estAR}.
\item [Step 2.] Compute the reduced-rank covariance estimator $\hat{\Sigma}_{Z}$ using the results in Proposition \ref{reduceRankDecomEstProp} based on the residuals from the autoregression conditioned on the AR coefficient estimates $\hat{\alpha}$.
\end{itemize}

~\\{\bf The model fitting procedure for the second scenario}.\\
Where there exist zero constraints on the AR coefficients (scenario 2), \eqref{sVAR_estAR} shows that the estimation of the AR coefficients $\alpha$ is confounded with the reduced-rank estimation of the noise covariance matrix $\Sigma_{Z}$. Therefore the reduced-rank covariance estimator is applied to a VAR model using the following iterative procedure.
\begin{itemize}
\item Start with initial estimators $\hat{\alpha}^{(0)}$ and $\hat{\Sigma}_{Z}^{(0)}$. 
\item Assume that at the $r$th iteration, the current estimators are $\hat{\alpha}^{(r)}$ and $\hat{\Sigma}_{Z}^{(r)}$, respectively.
Repeat the following steps 1 and 2 until convergence.
\begin{itemize}
\item[Step 1.] Compute $\hat{\alpha}^{(r+1)}$ according to \eqref{sVAR_estAR} by replacing $\hat{\Sigma}_{Z}$ with the current reduced-rank covariance estimator $\hat{\Sigma}_{Z}^{(r)}$.
\item[Step 2.] Compute $\hat{\Sigma}_{Z}^{(r+1)}$ by applying the results of Proposition \ref{reduceRankDecomEstProp} based on the residuals from the autoregression conditioned on the current constrained AR coefficient estimates $\hat{\alpha}^{(r+1)}$.
\end{itemize} 
\end{itemize}

~\\{\bf A latent space interpretation}.\\
We conclude this section by introducing a {\em latent space} setup that facilitates understanding and interpretation of a reduced-rank covariance VAR model. In particular, this latent space setup is useful in exploring {\em contemporaneous } dependence structure of the VAR process $\{Y_{t}\}$, which describes how synchronous values of different marginal series of $\{Y_{t}\}$ impact each other,  see e.g.,  \citet{Tunnicliffe2001,Demiralp2003, Moneta2004}. For $i=1,\ldots,K$, let $u_{i}\defeq(U_{i,1},\ldots, U_{i,d})^{'}$ be the $i$th row of the $K\times d$ matrix $U$ in \eqref{reduceRankCovMatr}. Then for two different marginal series of $\{Y_{t}\}$, say $\{Y_{t,i}\}$ and $\{Y_{t,j}\}$ ($i\ne j$), we have
\begin{eqnarray}\label{contempQuad}
\mathrm{cov}(Y_{t,i}, Y_{t,j}|~Y_{t-s}, 1 \le s\le p) = u_{i}^{'}\Lambda u_{j}.
\end{eqnarray}
The relation \eqref{contempQuad} shows that the conditional contemporaneous covariance between two different marginal series of $\{Y_{t}\}$ is represented by a weighted inner-product of the corresponding rows of $U$. To help interpret \eqref{contempQuad}, we postulate the existence of a $d$-dimensional Euclidean space of unobserved (latent) characteristics. The latent  characteristics determine the contemporaneous dependence between the marginal series of $\{Y_{t}\}$. We further assume that each marginal series of $\{Y_{t}\}$ is associated with a position in this latent space and the pattern of contemporaneous dependence among the $K$ marginal series of $\{Y_{t}\}$ can be characterized by their latent positions. Such a setup is also used in latent space network models, see e.g., \citet{Hoff2002, Hoff2005}. From \eqref{contempQuad} we can see that, when the above latent space setup is adopted to the reduced-rank covariance model \eqref{reduceRankCovMatr}, the $d$ dimensions of the latent space are represented by the columns of $U$ while the $K$ latent positions are given by the rows of $U$. Therefore the matrix $U$ provides a tool to represent the $K$-dimensional contemporaneous dependence structure in a lower-dimensional space. In addition, if we are able to find interpretations for different columns of $U$ by taking advantage of exogenous information, such interpretations will help identify the unobserved characteristics that are important in forming the contemporaneous dependence relationship. The heuristics behind such a latent space setup is similar to that of multidimensional scaling (MDS), see e.g., \citet{Borg1997}, in that both methods are concerned with ``spatial" representations of observed patterns of dependence among a group of subjects, such as the $K$ marginal series of $\{Y_{t}\}$ in our case. However, the MDS method is not model-based and it constructs spatial representations in an ad-hoc manner; in contrast, the above latent space setup leads to model-based graphical representations of the contemporaneous dependence structure via inference of the reduced-rank covariance model. In Section \ref{section_realExample_rr}, we illustrate via real data examples the use of this latent space setup in interpreting results from the reduced-rank covariance estimator in a VAR model.

\section{Numerical results}  
\subsection{Simulation}\label{section_simulation_rr} 
As mentioned in Section \ref{section_introduction_rr}, there are three major classes of covariance estimators under large dimensionality: shrinkage, regularization and structural covariance. The reduced-rank (RR) estimator can be viewed as a structure covariance estimator, as discussed in Section \ref{section_RR_independent_rr}. One difference between the three classes of covariance estimators is that, 
under finite samples, invertibility of the covariance estimator holds for the shrinkage and the structural approach, but not guaranteed for the regularization method. Due to this difference, in the simulation study we compare the reduced-rank covariance estimator with shrinkage estimators for their performance of estimating large dimensional covariance matrices from independent observations. The earliest attempt of shrinkage covariance estimation is given in \citet{Stein1975} and since then many shrinkage estimators have been proposed, see e.g., \citet{Dey1985, Daniels2001, Ledoit2003, Ledoit2004, Schafer2005}. A shrinkage covariance estimator is obtained by shrinking the sample covariance matrix towards a target covariance structure. The balance between these two extremes is controlled by the {\em shrinkage intensity}, a tuning parameter that needs to be estimated from data. A review of commonly-used target covariance structures is given in \citet{Schafer2005}.

We consider two shrinkage covariance estimators: one is proposed in \citet{Ledoit2004} (LW2004) and the other one is given by \citet{Schafer2005} (SS2005). The two shrinkage estimators differ in their choices of the target covariance structure. We generate independent replicates from a $K$-dimensional Gaussian $N(0, \Sigma_{Z})$ under three cases:
\begin{itemize} 
\item[(\MakeUppercase{\romannumeral 1})] $\Sigma_{Z} = I_{K}$.
\item[(\MakeUppercase{\romannumeral 2})] $\Sigma_{Z}$ has all covariances set to 0.16 and variances set to $\{1.0, 1.0, 0.5,\ldots,0.5\}$ (the first two entries are 1.0 and the remaining entries are 0.5).
\item[(\MakeUppercase{\romannumeral 3})] $\Sigma_{Z}$ has the $(i,j)$th $(i \ne j)$ covariance set to $(-1)^{(i+j)}\times0.10$ and variances set to $\{0.47,0.49,\ldots,0.73,0.75\}$ (the ascending sequence from 0.47 to 0.75 with increment 0.02).
\end{itemize}

\noindent Case (\MakeUppercase{\romannumeral 1}) gives a very simple covariance structure; Case (\MakeUppercase{\romannumeral 2}) serves as an example of the reduced-rank covariance structure \eqref{reduceRankCovMatr} with the reduced-rank $d=3$; Case (\MakeUppercase{\romannumeral 3}) does not satisfy the reduced-rank covariance model \eqref{reduceRankCovMatr}. We take the dimension $K=15$ and the sample size $T=50, 100, 200, 400$.  In applying the RR covariance estimator, the reduced-rank $d$ is selected from $\{1,\ldots,14\}$ according to a minimum BIC, which is computed as in \eqref{rr_BIC}. In applying the two shrinkage estimators LW2004 and SS2005, their shrinkage intensities are determined analytically as described in \citet{Ledoit2004} and \citet{Schafer2005}, respectively. 

First we investigate the RR covariance estimator's performance of inferring the reduced-rank $d$ when the true underlying covariance matrix admits a reduced-rank structure \eqref{latentVarModel}. We use the $\Sigma_{Z}$ in Case (\MakeUppercase{\romannumeral 2}) as an example, which satisfies the reduced-rank covariance assumption with the reduced-rank $d=3$. Table \ref{sim1_distEstReducedRank_case3} summarizes the frequencies (out of 500 replications) of the estimated reduced-rank $\hat{d}$ for different sample sizes. We can see that when the sample size is relatively small, e.g., $T=50$ and $100$, the RR covariance estimator tends to under-estimate the reduced-rank; as the sample size $T$ increases, the probability of selecting the correct reduced-rank increases accordingly. In particular, when the sample size $T$ reaches 400, the RR covariance estimator has a large probability of selecting the correct reduced-rank $d=3$.
 
\begin{table}[hbt]
\begin{center}
\begin{tabular}{r|r|r|r|r}
\hline
$T$       & $\hat{d}=1$ & $\hat{d}=2$ & $\hat{d}=3$ & $\hat{d}\ge 4$ \\
\hline
50        & 447          & 52           &  1          & 0  \\
\hline
100      & 304          & 153         &  43         & 0  \\
\hline
200      & 30           & 146          &  324       & 0  \\
\hline
400      & 0             & 0            & 500         & 0   \\
\hline
\end{tabular}
\end{center}
\caption{Frequencies of the estimated reduced-rank $\hat{d}$ of the RR covariance estimator for Case (\MakeUppercase{\romannumeral 2}). The true reduced-rank $d=3$ and results are based on 500 replications.}
\label{sim1_distEstReducedRank_case3}
\end{table}

Next we compare the performance of the RR covariance estimator with the two shrinkage covariance estimators LW2004 and SS2005. We use two metrics for the comparison: the first metric is based on {\em Stein's loss} (SL) \citet{James1961}, which is defined by $\mathrm{SL}(\hat{\Sigma}_{Z})\defeq\mathrm{tr}(\hat{\Sigma}_{Z}\Sigma_{Z}^{-1}) - \log|\hat{\Sigma}_{Z}\Sigma_{Z}^{-1}| - K$. It can be shown that Stein's loss $\mathrm{SL}(\hat{\Sigma}_{Z})$ is equal to (up to a constant multiplier) the {\em Kullback-Leibler divergence} \citet{Kullback1951} between two $K$-dimensional Gaussians $N(0, \hat{\Sigma}_{Z})$ and $N(0, \Sigma_{Z})$; and the second metric is the mean squared error (MSE), which is defined by $\mathrm{MSE}(\hat{\Sigma}_{Z})\defeq ||\hat{\Sigma}_{Z} - \Sigma_{Z}||_{2}^2$. We use Stein's loss to characterize the eigen-structure of covariance estimators while we also consider point-wise estimation accuracy of covariance estimators by comparing their MSE.

Table \ref{sim1_reductionFromSamcov_SLandMSE} summarizes the percentage reductions (with standard errors in brackets) in Stein's loss and MSE of each covariance estimator as compared to the sample covariance matrix. For each setting, the largest reduction among the three estimators is marked in bold. We can see that all three covariance estimators lead to improvement over the sample covariance matrix for both Stein's loss and MSE. For Case (\MakeUppercase{\romannumeral 1}), where the true $\Sigma_{Z}=I_{K}$ has a very simple structure, all three covariance estimators achieve similar improvement over the sample covariance matrix for both Stein's loss and MSE. It is more interesting to compare the three covariance estimators when the structure of $\Sigma_{Z}$ becomes more complicated in Cases (\MakeUppercase{\romannumeral 2}) and (\MakeUppercase{\romannumeral 3}). For Case (\MakeUppercase{\romannumeral 2}), where the reduced-rank covariance assumption \eqref{latentVarModel} is satisfied, we can see that the RR covariance estimator leads to significant improvement over the sample covariance matrix in Stein's loss for various sample sizes. At the same time, for small-to-medium sample sizes, such as $T=50$ and $100$, the improvement in Stein's loss from the two shrinkage estimators LW2004 and SS2005 is comparable to that from the RR covariance estimator; as the sample size increases, such as $T=200$ and $400$, the improvement in Stein's loss from the two shrinkage estimators becomes much less significant. We can also see that the improvement in MSE from all three covariance estimators is less significant as compared to their improvement in Stein's loss.  For Case (\MakeUppercase{\romannumeral 3}), it is interesting to see that even if $\Sigma_{Z}$ does not satisfy the reduced-rank covariance model \eqref{latentVarModel}, the RR covariance estimator still results in significant improvement in Stein's loss over the sample covariance matrix for all sample sizes. In addition, the improvement in Stein's loss from both the RR covariance estimator and the two shrinkage estimators is much more significant than their improvement in MSE. To explain the performance of the RR covariance estimator in Case (\MakeUppercase{\romannumeral 3}), we point out that the largest eigen-value of $\Sigma_{Z}$ in Case (\MakeUppercase{\romannumeral 3}) is dominant over the remaining eigen-values. As a result, the eigen-structure of $\Sigma_{Z}$ is close to that of a reduced-rank covariance matrix, even though $\Sigma_{Z}$ in Case (\MakeUppercase{\romannumeral 3}) does not satisfy the reduced-rank covariance model \eqref{reduceRankCovMatr}.

\begin{landscape}
\begin{table}[hbt]
\begin{center}
\begin{tabular}{l|r|c|c|c|c|c|c} 
\hline
\multicolumn{2}{c|}{~}  & \multicolumn{3}{c|}{percentage reduction in SL} & \multicolumn{3}{c}{percentage reduction in MSE} \\
\hline
$\Sigma_{Z}$                                      & $T$  &\quad RR\quad\quad  & LW2004  & SS2005 & \quad RR \quad\quad & LW2004  & SS2005 \\
\hline
\MakeUppercase{\romannumeral 1}       & 50    &  {\bf 99.1} (0.053)  & 98.3 (0.080)   & 97.8 (0.096) &  {\bf 99.0} (0.067) & 98.1 (0.094)  & 97.5 (0.111) \\
                                                          & 100  &  {\bf 99.2} (0.051) & 98.5 (0.076)   & 97.9  (0.100) &  {\bf 99.1} (0.056) & 98.4 (0.082)  & 97.7 (0.105) \\
                                                          & 200  &  {\bf 99.2} (0.055) & 98.6  (0.074)  & 97.8  (0.101) &  {\bf 99.1} (0.060) & 98.5 (0.079)  & 97.8 (0.106) \\
										    & 400  &  {\bf 99.2}  (0.045) & 98.6  (0.073)  & 97.7  (0.109) &  {\bf 99.2} (0.047) & 98.5 (0.074)  & 97.6 (0.110) \\
\hline
\MakeUppercase{\romannumeral 2}     & 50    &  {\bf 68.3} (0.242)  & 50.1 (0.219)    & 47.4 (0.226)  &  18.3          (0.460) & 12.4 (1.243)  & {\bf 14.8}  (1.228) \\
                                                       & 100   &  {\bf 48.7} (0.468)  & 30.3 (0.162)    & 27.5 (0.139)  &  ~0.0         (0.531) & ~6.5 (1.067)  & ~{\bf 8.6} (1.017) \\
                                                       & 200   &  {\bf 51.2} (0.927)  & 16.3 (0.102)    & 14.5 (0.079)  &  ~{\bf 7.3} (1.056) & ~2.7 (0.839)  & ~4.0         (0.799) \\
										  & 400  &  {\bf 64.3} (0.277)  & ~8.6 (0.073)    & ~7.5 (0.056)  &  {\bf 22.9} (0.298) & ~1.5 (0.612)  & ~2.2         (0.580) \\
\hline
\MakeUppercase{\romannumeral 3}     & 50    &  {\bf 77.8} (0.334) & 69.7 (0.204)  & 68.1 (0.242) & 37.2        (1.562) &  38.5 (0.880) & {\bf 39.4} (0.888)\\
                                                       & 100   &  {\bf 71.4} (0.254) & 50.5 (0.183) & 47.2 (0.264) & {\bf 47.6} (0.601) &  23.0 (0.911) & 23.9         (0.878)\\
                                                       & 200   &  {\bf 53.9} (0.275) & 31.8 (0.128) & 28.7 (0.196) & {\bf 37.5} (0.381) &  12.5 (0.813) & 14.0         (0.756)\\
										  & 400  &   {\bf 20.5} (0.458) & 18.0 (0.097) & 15.7 (0.163) & {\bf 16.3} (0.348) &  ~7.7 (0.633) & ~8.6        (0.571)\\
\hline
\end{tabular}
\end{center}   
\caption{Percentage reductions (with standard errors in brackets) in Stein's loss (SL) and MSE of the RR, the LW2004 and the SS2005 covariance estimators as compared to the sample covariance matrix. Results are based on 500 replications.}
\label{sim1_reductionFromSamcov_SLandMSE}
\end{table} 
\end{landscape}

\subsection{Real data examples}\label{section_realExample_rr}
We apply the reduced-rank covariance estimator to VAR modeling of two real data examples. The first example is concerned with stock returns in S\&P 500 and corresponds to the first scenario in Section \ref{section_RRappToVAR_rr}, i.e., there are no constraints on the AR coefficients of the VAR model. The second example is a time series of temperatures in southeast China and corresponds to the second scenario, i.e., there are zero-constraints on the AR coefficients. For both examples, we use the latent space setup introduced in Section \ref{section_RRappToVAR_rr} to interpret results of the reduced-rank covariance estimation.

\begin{figure}[p]
\begin{center}
\includegraphics[width=0.95\textwidth,height=0.90\textwidth]{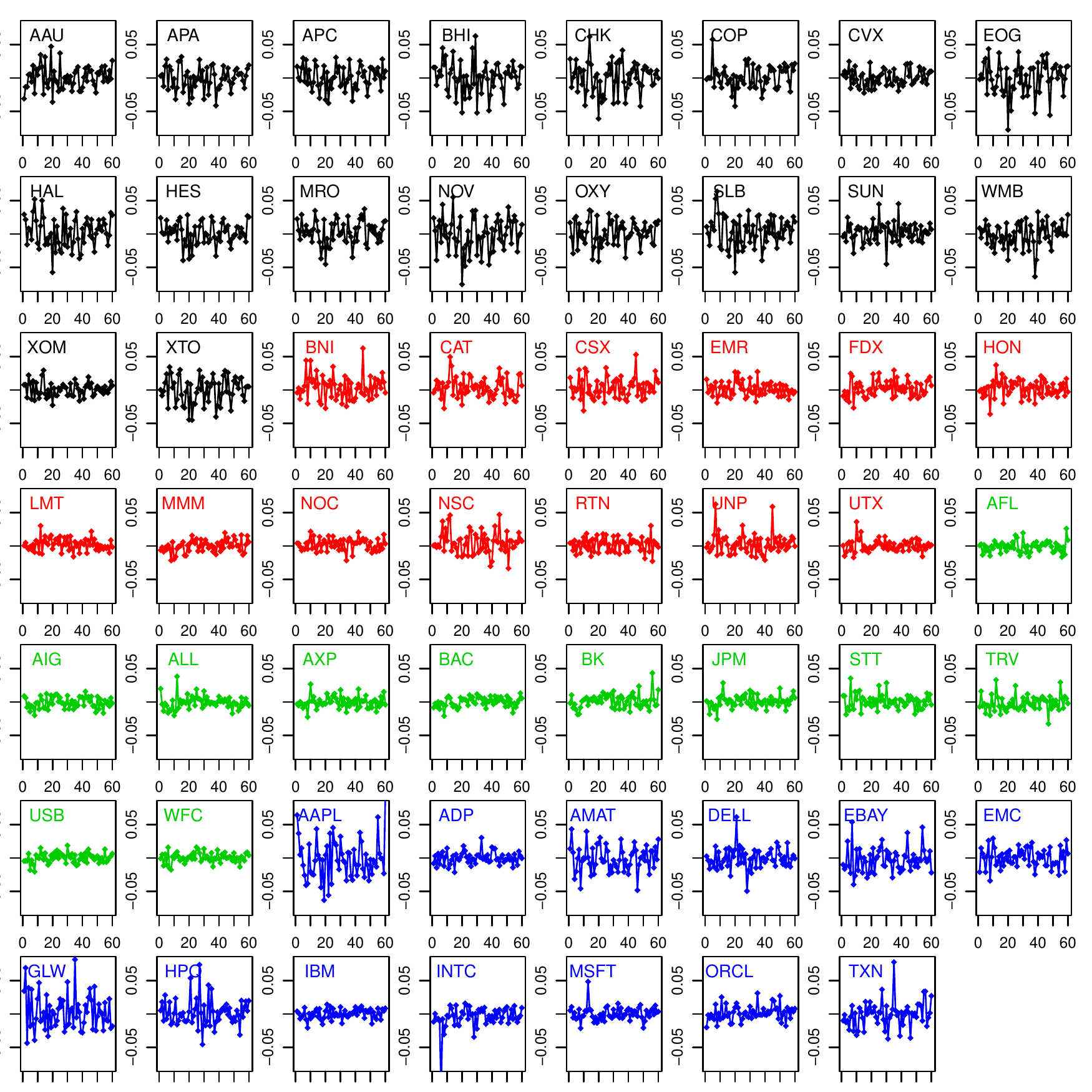}
\caption{The first 60 observations of the return series. The color indicates the sector each stock belongs to: {\em energy} (black), {\em industry} (red), {\em finance} (green), {\em technology} (blue).}
\label{spStock_rr_2006_data}
\end{center}
\end{figure}

~\\{\bf Stock returns from S\&P 500}. In the first example, the data consist of daily returns of $K=55$ stocks in S\&P 500 and the stocks come from 4 sectors: {\em energy}, {\em industry}, {\em finance} and {\em technology}. The returns are calculated as the logarithm of the ratio between two consecutive daily closing prices from the $T=252$ trading days in 2006. Figure \ref{spStock_rr_2006_data} displays the first 60 observations of the return series.

\begin{figure}[p]
\begin{center}$ 
\begin{array}{cc}
\subfloat[the BIC curve]{\includegraphics[width=0.5\textwidth]{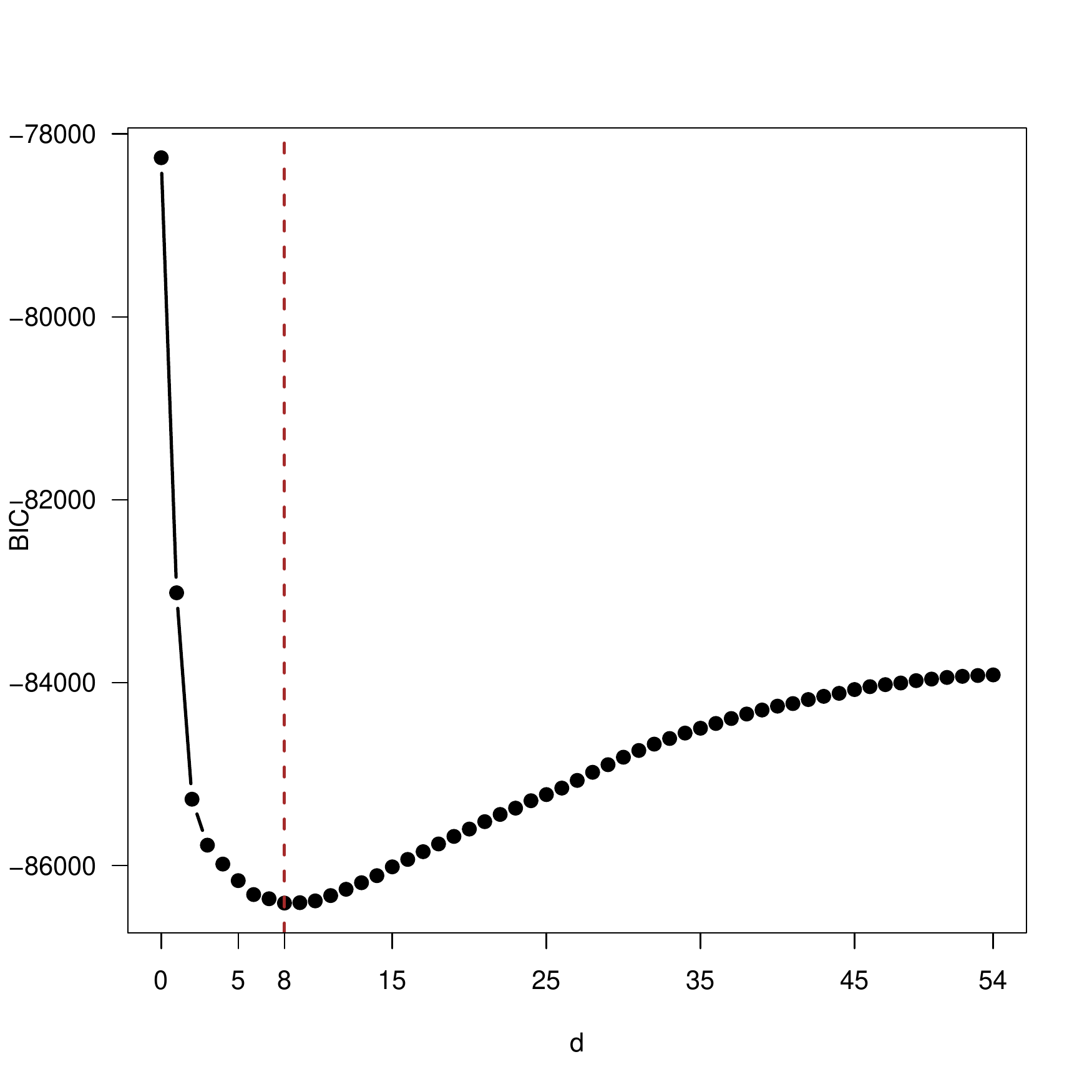}} &
\subfloat[dimension 1 vs dimension 2]{\includegraphics[width=0.5\textwidth]{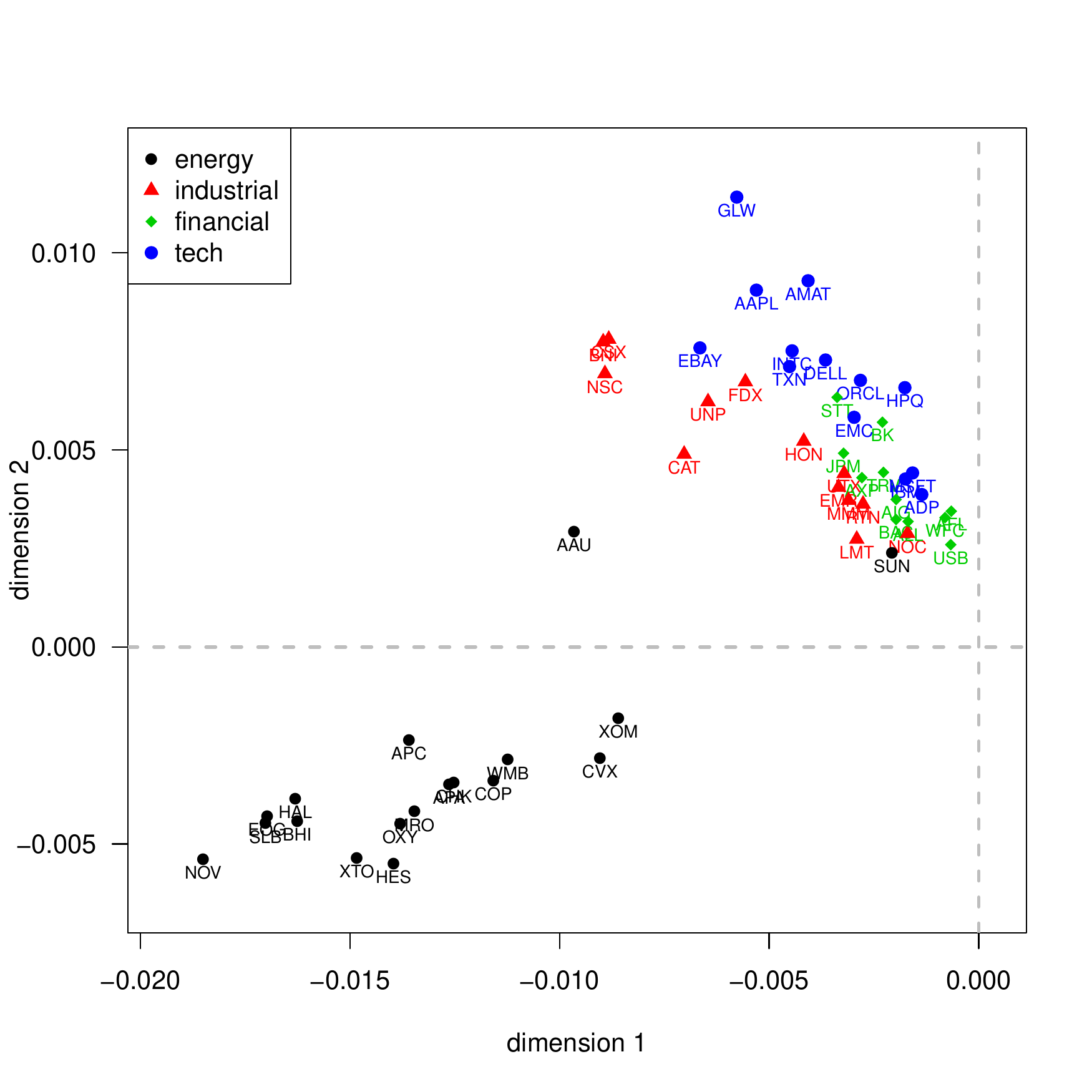}}  \\
\subfloat[dimension 1 vs dimension 3]{\includegraphics[width=0.5\textwidth]{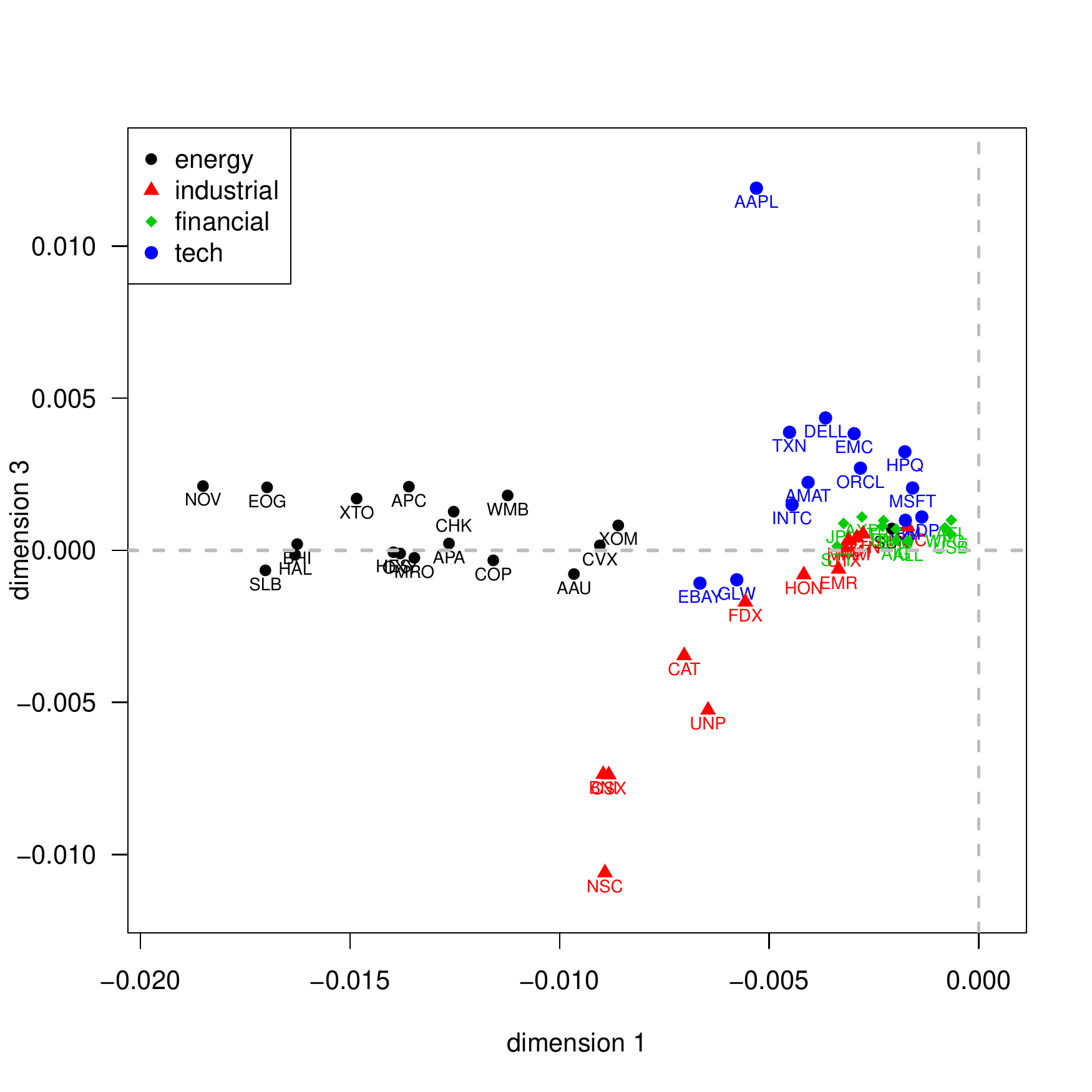}}  & 
\subfloat[dimension 2 vs dimension 3]{\includegraphics[width=0.5\textwidth]{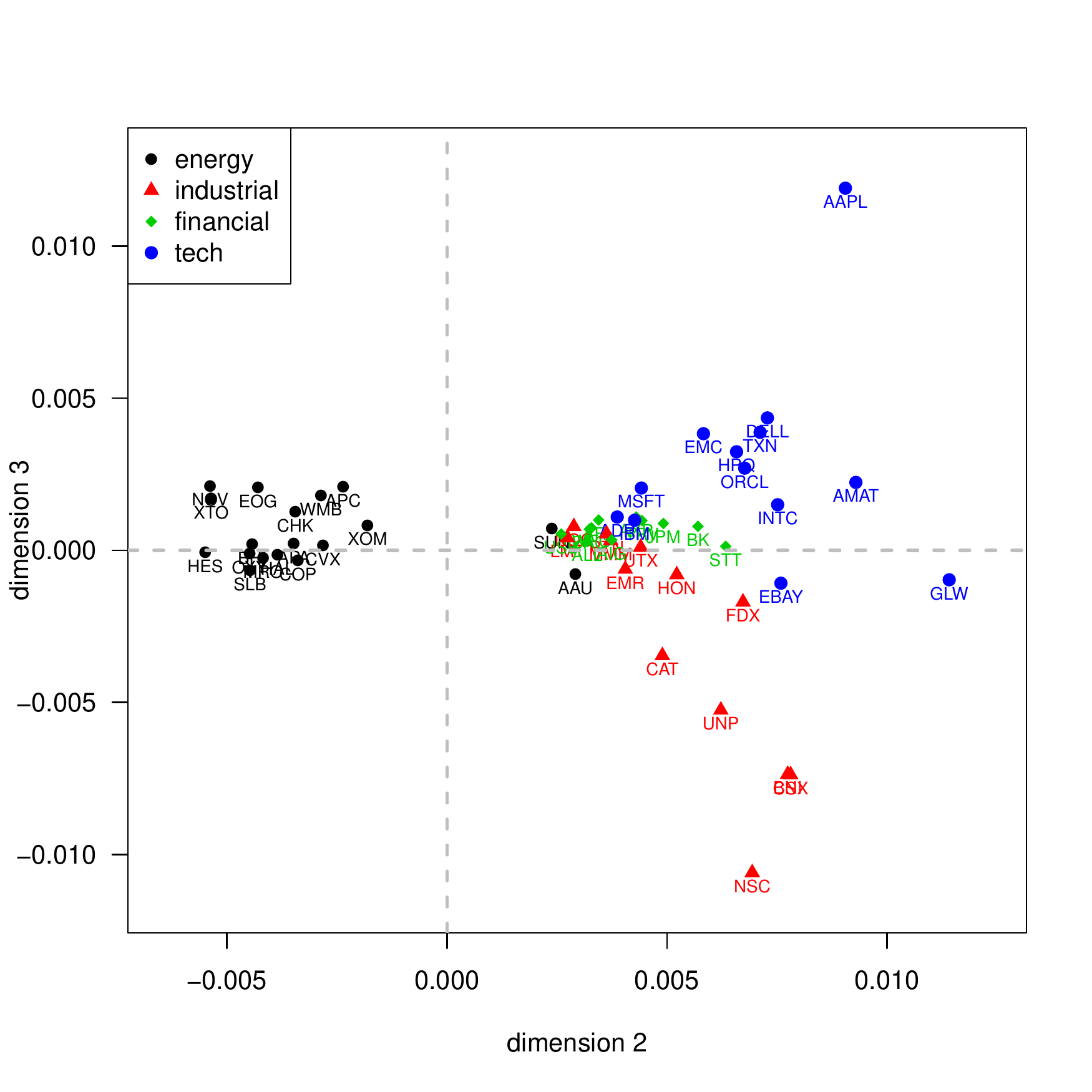}} 
\end{array}$
\caption{Panel (a): The BIC curve as the reduced-rank $d$ varies from 1 to 54. Panels (b), (c) and (d): Layouts of the 55 stocks in the first 3 dimensions of the latent space. The color indicates the sector each stock belongs to: {\em energy} (black), {\em industry} (red), {\em finance} (green), {\em technology} (blue).}
\label{spStock_rr_combine}
\end{center}
\end{figure}

Our interest is to describe the pattern of contemporaneous dependence between the returns of the 55 stocks. For this purpose, we apply the reduced-rank covariance estimator to the VAR modeling of the 55-dimensional return series. We first use the 2-step method, which is described in the first scenario in Section \ref{section_RRappToVAR_rr}, to fit a VAR model with unconstrained AR coefficients and a reduced-rank noise covariance matrix. In particular, we first fit an unconstrained VAR(1) model to the 55-dimensional return series, where the autoregression order $1$ is selected from $\{0,1,2,3\}$ according to a minimum BIC. Then we obtain the reduced-rank covariance estimator based on the residuals from the fitted autoregression. We select the reduced-rank $d$ from $\{1,2,\ldots,54\}$ according to a minimum BIC, which is computed in equation \eqref{rr_BIC}. Panel (a) in Figure \ref{spStock_rr_combine} displays the BIC curve as $d$ varies and it shows that the minimum BIC occurs when $d=8$. In other words, the contemporaneous dependence structure between the 55 stocks' returns can be well represented in a 8-dimensional latent space. Panels (b), (c) and (d) in Figure \ref{spStock_rr_combine} display the layouts of the 55 stocks in the first 3 dimensions of the 8-dimensional latent space, where the color indicates the sector each stock belongs to. Panel (b) corresponds to the first 2 dimensions of the latent space and we can observe a ``clustering" phenomenon of the 55 stocks in these 2 dimensions. Specifically, the within-sector contemporaneous dependence is most noticeable among the {\em energy} stocks, since they are positioned close to each other while far away from the origin of the latent space. We also observe that most of the {\em energy} stocks have the opposite sign along the second dimension of the latent space as compared to stocks from the {\em industry}, {\em finance} and {\em technology} sectors. This means that returns of the {\em energy} stocks are negatively contemporaneously related to stock returns from the other 3 sectors. On the other hand, the within-sector contemporaneous dependence is much weaker among the {\em finance} stocks, since those stocks are positioned close to the origin of the latent space. Moreover, panel (b) also shows that the first 2 dimensions provide information for separating the {\em energy} sector from the other 3 sectors, but not for distinguishing among the {\em industry}, {\em finance} and {\em technology} stocks. One exception is that there also exists separation between the {\em industry} and the {\em technology} sectors. This separation becomes more noticeable after we take into account the third dimension of the latent space. From panels (c) and (d), both of which display the third dimension along the vertical direction, we can see that the third dimension is informative for separating the {\em industry} from the {\em technology} stocks, while it has little power for distinguishing between the {\em energy} and the {\em finance} sectors. 

\begin{figure}[p]
\begin{center}
\includegraphics[width=0.85\textwidth,height=0.85\textwidth]{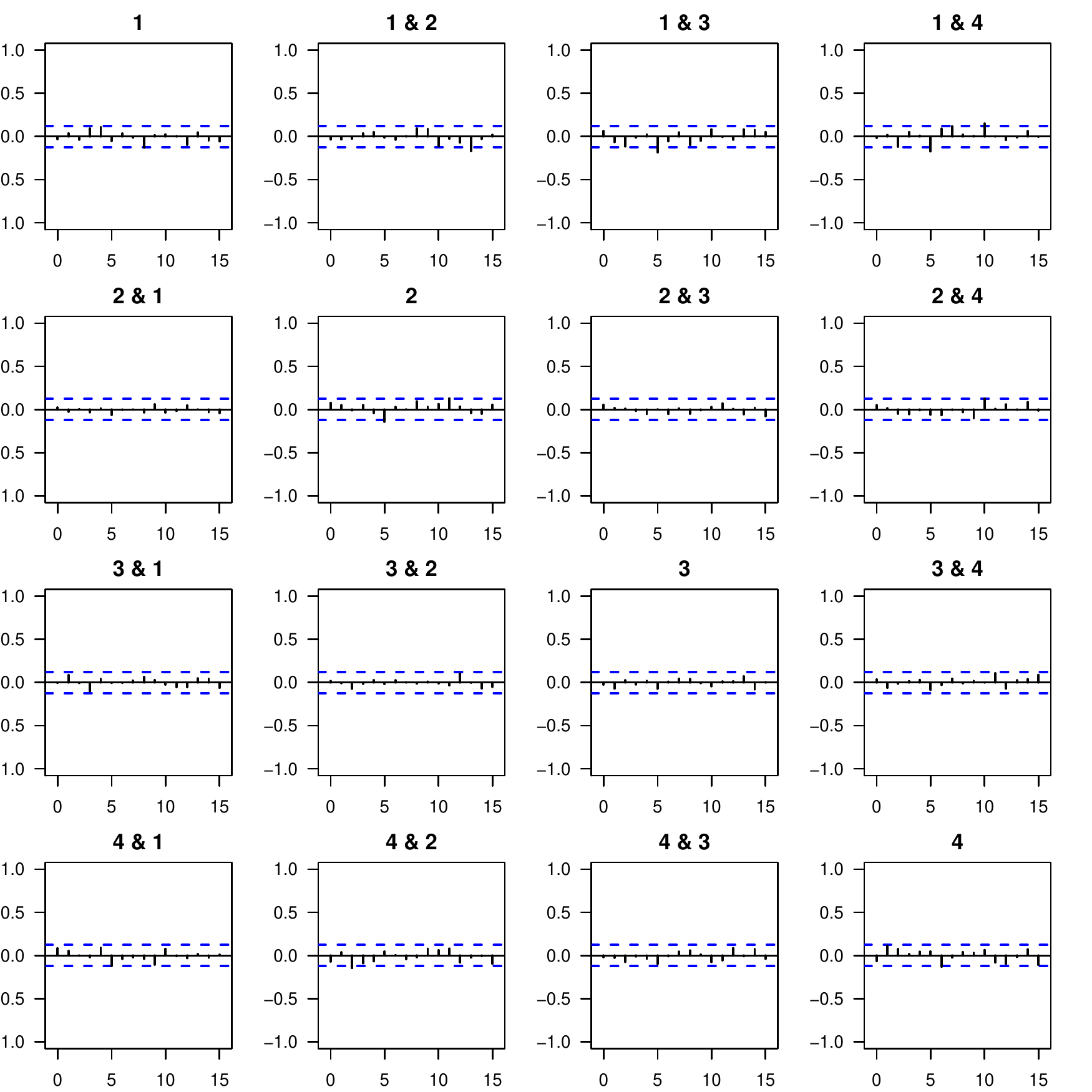}
\caption{The ACF and CCF plots among the first 4 dimensions of the estimated latent variable $\hat{\delta}_{t}$.}
\label{spStock_rr_ACF_CCF_delta}
\end{center}
\end{figure}

As a diagnostic check, Figure \ref{spStock_rr_ACF_CCF_delta} displays the auto-correlation (ACF) and cross-correlation functions (CCF) among the first 4 dimensions of the estimated latent variable $\hat{\delta}_{t}$ as computed in \eqref{reduceRankDecomEstDelta} and it exhibits little significant auto- or cross- correlation. In fact, we observe little significant auto- or cross- correlation among all 8 dimensions of $\hat{\delta}_{t}$. This observation is consistent with the assumptions of the reduced-rank covariance model.

\begin{figure}[p]
\begin{center}$ 
\begin{array}{cc}
\subfloat[confidence interval of AR estimates]{\includegraphics[width=0.5\textwidth]{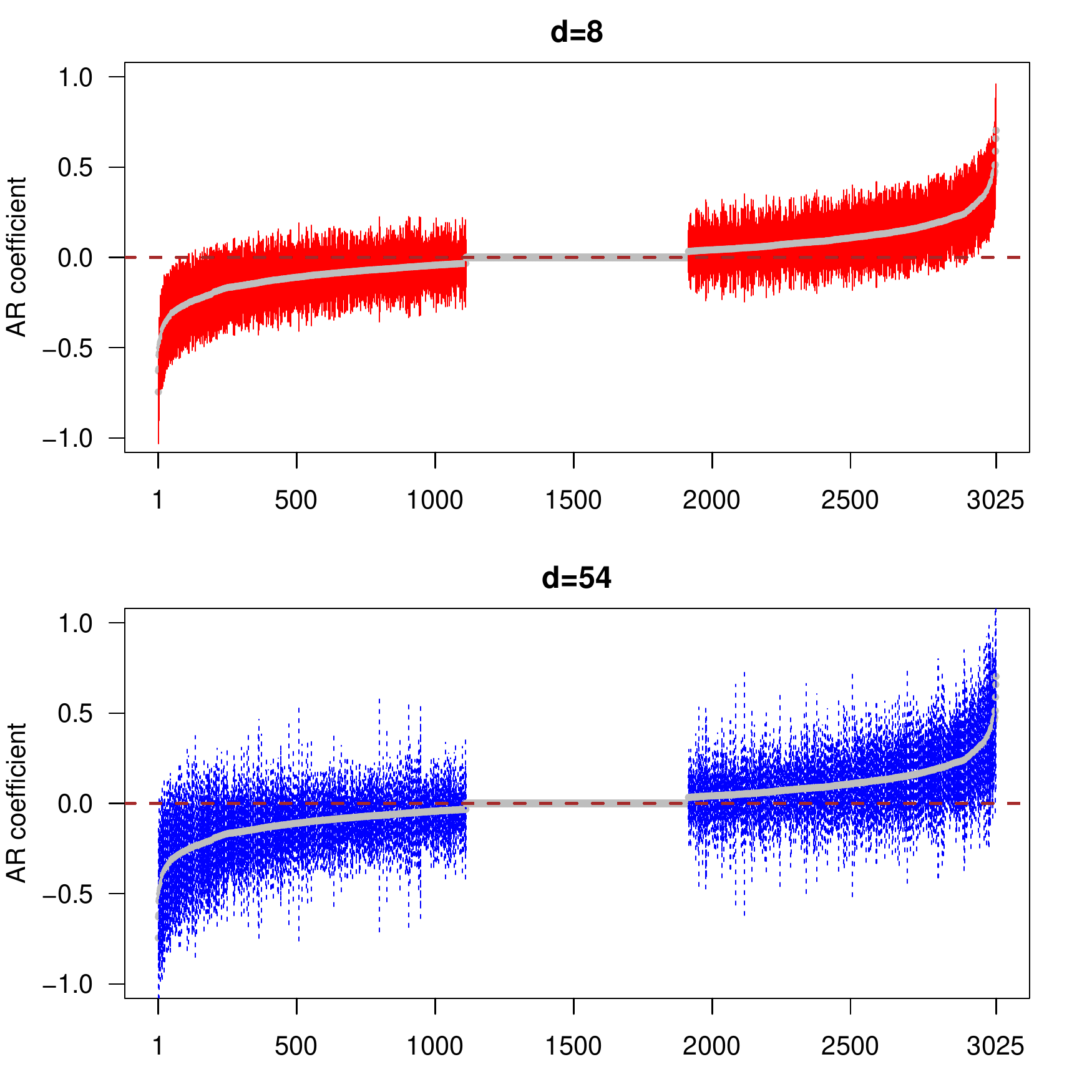}}  & 
\subfloat[approximate 1-step MSE]{\includegraphics[width=0.5\textwidth]{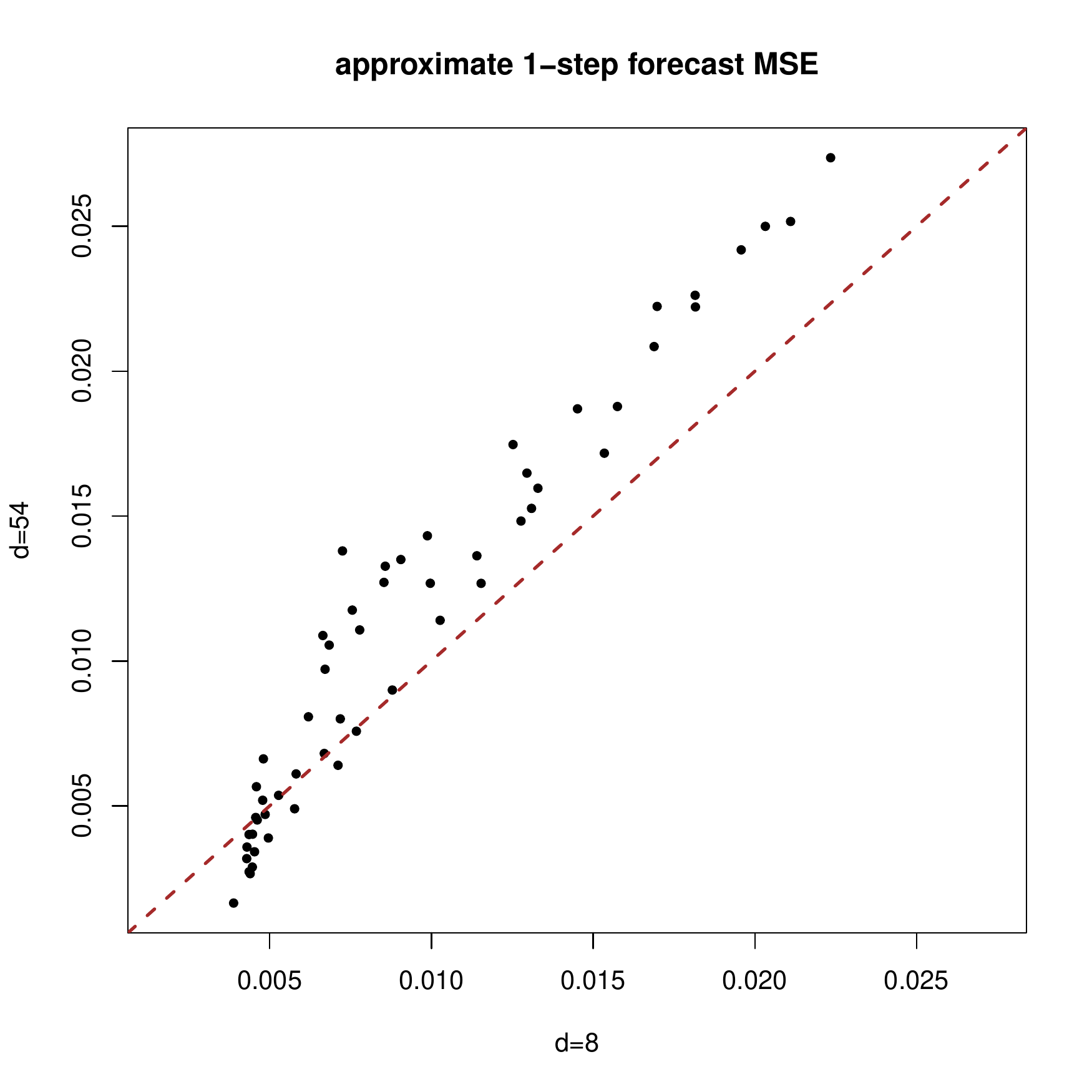}} 
\end{array}$
\caption{Panel (a): Comparison between the confidence intervals of the AR coefficient estimates in the two sparse VAR models with $d=8$ (top) and $d=54$ (bottom). The solid curve shows the AR coefficient estimates in ascending order. Each vertical line indicates $\pm$1.96 the corresponding standard error.. Panels (b): Comparison between the diagonal entries of the approximate 1-step forecast MSE of the two sparse VAR models when $d=8$ (x-axis) and $d=54$ (y-axis).}
\label{spStock_rr_full_reduced}
\end{center}
\end{figure}

Applying the reduced-rank covariance estimator to large-dimensional VAR modeling might also lead to improvement over the scenario where an unrestricted covariance estimator is used for estimating the noise covariance matrix. Here the unrestricted covariance estimator refers to the sample covariance matrix of the residuals from fitted autoregression and it corresponds to the case where $d=K-1$ in the reduced-rank covariance model \eqref{reduceRankCovMatr}. For the comparison between the reduced-rank and the unrestricted covariance estimators, first we apply the 2-stage approach introduced in \citet{Davis2012} to the 55-dimensional returns series to determine non-zero AR coefficient estimates. To reduce the computational effort, we take into account the above information regarding which AR coefficient estimates are non-zero and fix the reduced-rank $d=8$ and $d=54$, respectively, while we apply the second model fitting procedure in Section \ref{section_RRappToVAR_rr} to the returns series. We finally obtain two sparse VAR(1) models each with a reduced-rank $d=8$ and $d=54$, respectively. Even if the selection of non-zero AR coefficient estimates is identical between these two VAR models, the complexity of the noise covariance estimator will impact the VAR models in the following two aspects: the confidence intervals of AR coefficient estimates and the forecast mean squared error (MSE) will be different. Panel (a) in Figure \ref{spStock_rr_full_reduced} displays the confidence intervals of the AR coefficient estimates from the two sparse VAR models with $d=8$ and $d=54$, respectively. The solid curve shows the AR coefficient estimates in ascending order and each vertical line indicates $\pm$1.96 times the corresponding standard error. From panel (a) we can see that reducing the complexity of the noise covariance estimator from $d=54$ to $d=8$ results in narrower confidence intervals for AR coefficients. Such narrower confidence intervals help to identify significant temporal relationships in VAR models. We can also see that the confidence intervals in the case of the reduced-rank noise covariance estimator are more stable as compared to those in the VAR model with the unconstrained noise covariance estimator. Next we compare the forecast MSE of the two sparse VAR models. The MSE matrix of 1-step forecast of a VAR($p$) model with {\em estimated} AR coefficient matrices $\hat{A}_{1},\ldots,\hat{A}_{p}$ is defined as 
\begin{eqnarray} 
\mathrm{fMSE}(1) \defeq \mathbb{E}(Y_{t+1}-\displaystyle\sum_{k=1}^{p}\hat{A}_{k}\hat{Y}_{t+1-k})(Y_{t+1}-\displaystyle\sum_{k=1}^{p}\hat{A}_{k}\hat{Y}_{t+1-k})^{'}, \label{oneStepMSEdef}
\end{eqnarray}
where $\hat{Y}_{t-k}\defeq Y_{t}$ for $k\le 0$. Results in Appendix \ref{section_appendix_mse_rr} show that the forecast MSE matrix \eqref{oneStepMSEdef} can be approximated by the estimates of the AR coefficients $\hat{A}_{1},\ldots,\hat{A}_{p}$ and the noise covariance matrix $\hat{\Sigma}_{Z}$. Panel (b) in Figure \ref{spStock_rr_full_reduced} compares the diagonal entries of the approximate 1-step forecast MSE matrices between the two sparse VAR models with $d=8$ and $d=54$, respectively. We can see that the reduced-rank covariance estimator leads to smaller 1-step forecast MSE than the unrestricted covariance estimator.

~\\{\bf Temperatures in southeast China}. This example is concerned with the monthly temperature series of $K=7$ cities in southeast China \footnote{The seven cities are {\em Anqing, Dongtai, Hangzhou, Hefei, Huoshan, Nanjing} and {\em Shanghai}.} from January 1988 to December 1998 with $T=132$ observations, e.g., see \citet{Pan2008}.

\begin{figure}[p]
\begin{center}$ 
\begin{array}{cc}
\subfloat[the geo-space]{\includegraphics[width=0.5\textwidth]{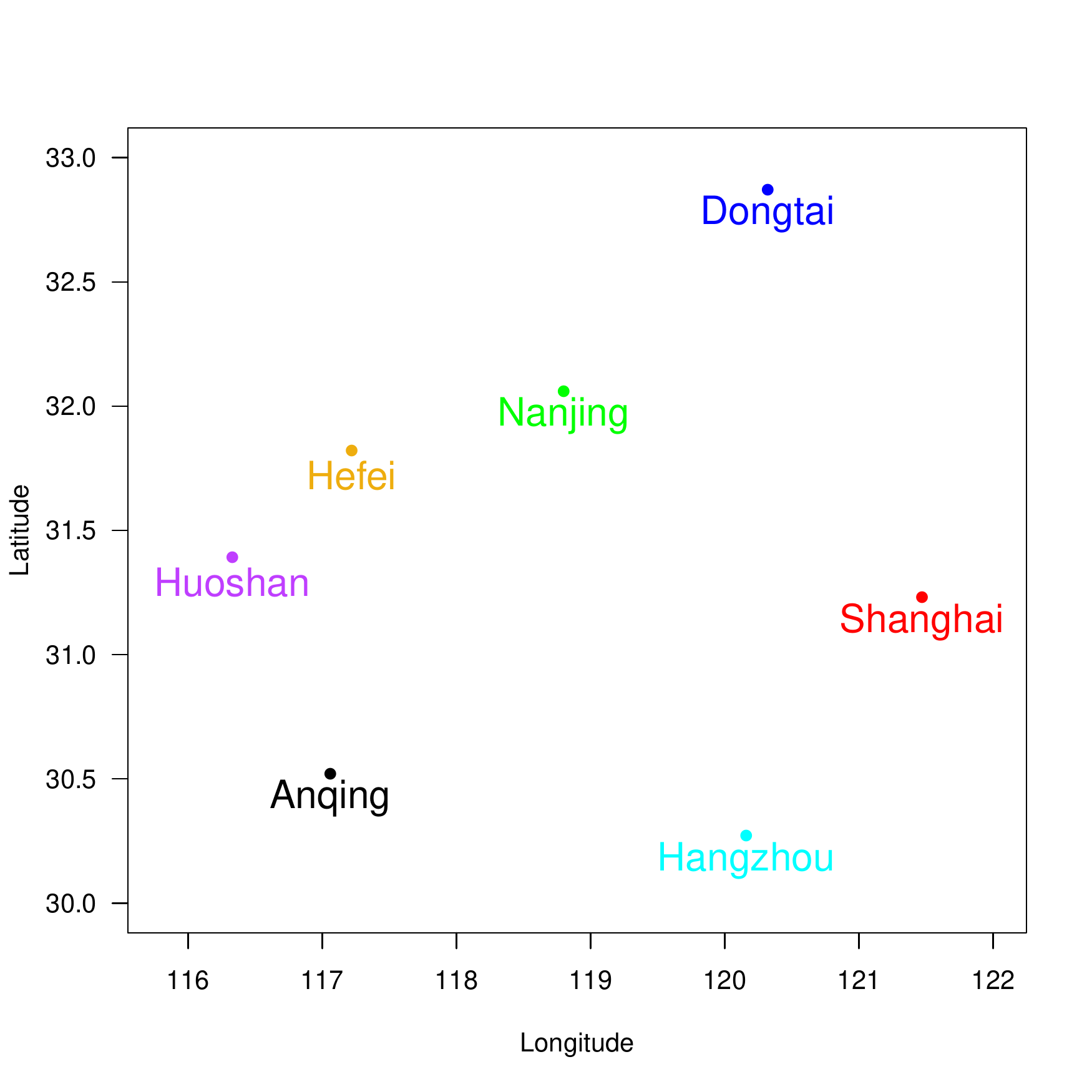}} &
\subfloat[the latent space]{\includegraphics[width=0.5\textwidth]{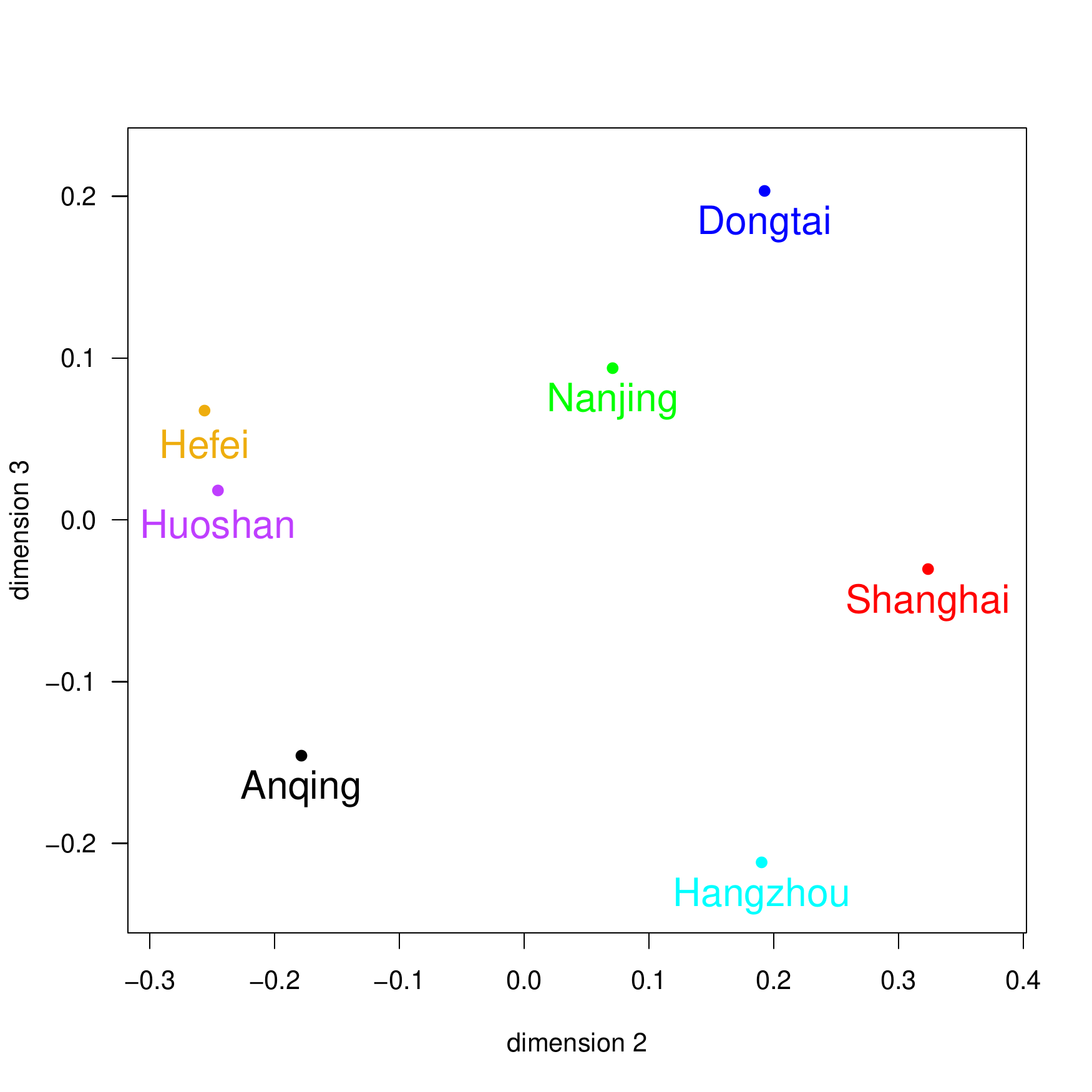}}  \\
\subfloat[pairwise distance]{\includegraphics[width=0.5\textwidth]{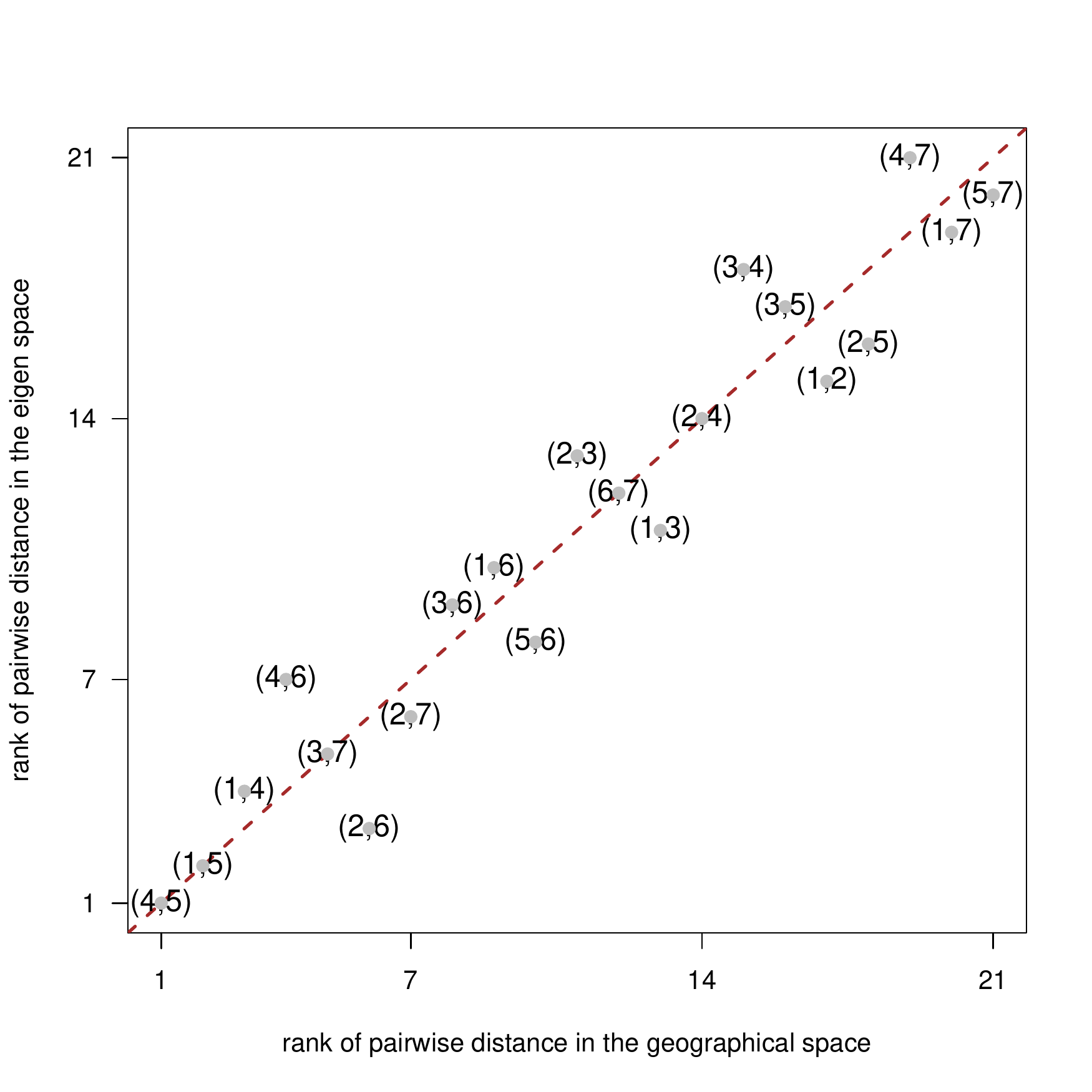}}  & 
\subfloat[ACF and CCF]{\includegraphics[width=0.5\textwidth]{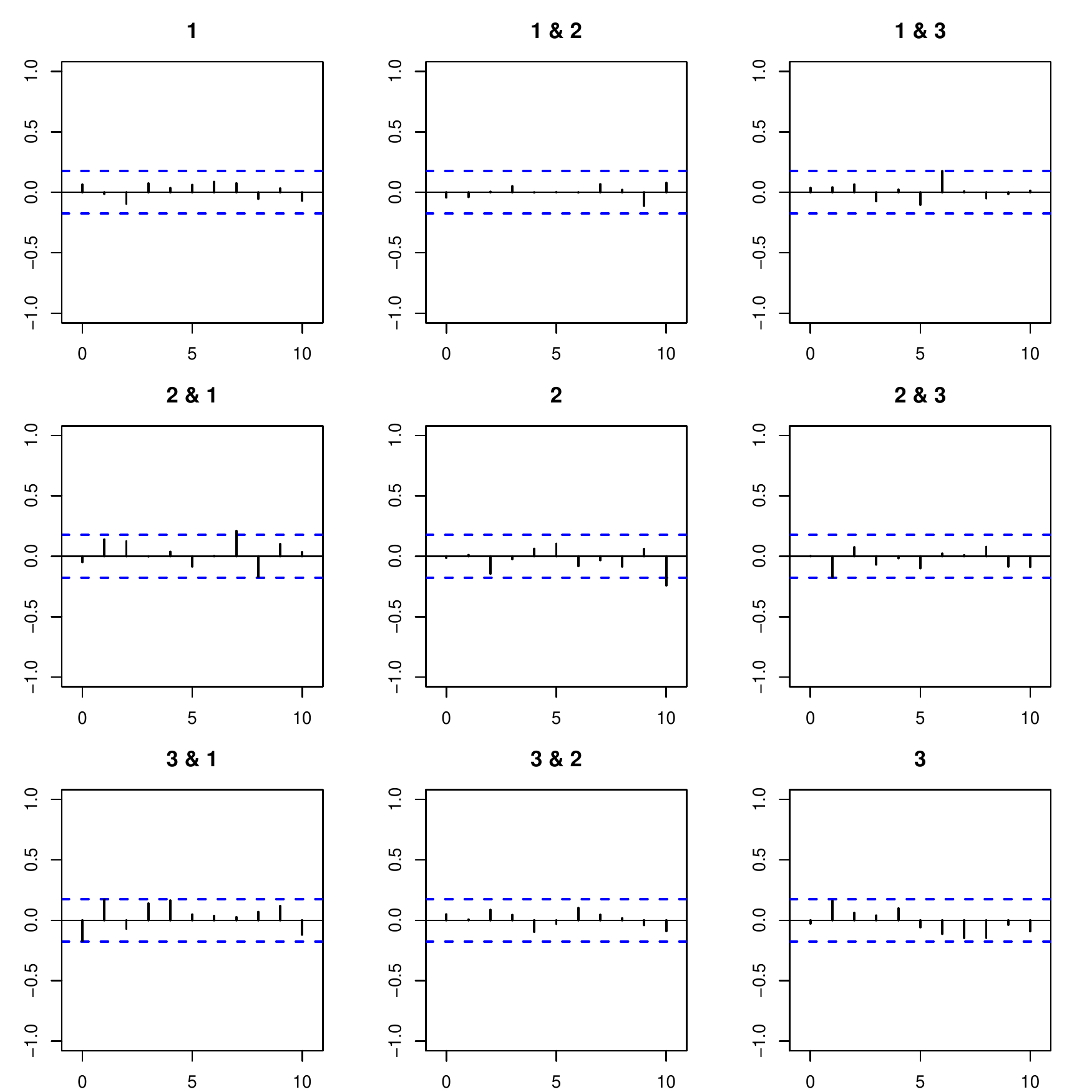}} 
\end{array}$
\caption{Panels (a): Actual geographical locations of the 7 cities. Panel (b): Estimated latent positions of the 7 cities. Panel (c): Ranks of pairwise distances among the 7 cities in the geographical space (x-axis) and in the latent space (y-axis). The numbers stand for: 1-Anqing, 2-Dongtai, 3-Hangzhou, 4-Hefei, 5-Huoshan, 6-Nanjing and 7-Shanghai. Panel (d): The ACF and CCF plots of the 3 dimensions of the estimated latent variable $\hat{\delta}_{t}$.}
\label{cityTemperature_rr_combine}
\end{center}
\end{figure}

We are interested in the contemporaneous dependence structure between the 7 cities' temperature movements. For this purpose, we apply the reduced-rank covariance estimation in the VAR modeling of the 7-dimensional temperature series. We use the iterative procedure, which corresponds to the second scenario in Section \ref{section_RRappToVAR_rr}), to fit a VAR model with sparse AR coefficients and a reduced-rank noise covariance matrix. Specifically, for each $d \in \{1,2,\ldots,6\}$, we use the 2-stage approach introduced in \citet{Davis2012} to determine zero constraints on the AR coefficients according to a minimum BIC. In applying the 2-stage approach, the order of autoregression $p$ is selected from $\{0,1,\ldots,8\}$. Then we choose the reduced-rank $d$ from $\{1,2,\ldots,6\}$ according to a minimum BIC as well. We finally obtain a VAR(1) model with 29 non-zero AR coefficients and reduced-rank $d=3$. The selection of $d=3$ suggests that the core structure of contemporaneous dependence between the 7 cities' temperatures can be represented in a $3$-dimensional latent space. To obtain insight about this $3$-dimensional latent space, we compare the 7 cities' actual geographical locations with their positions in the estimated latent space. The findings are summarized in Figure \ref{cityTemperature_rr_combine}. Panel (a) in the figure displays the 7 cities'  geographical locations (longitude vs latitude) while panel (b) shows their estimated latent positions (dimension 2 vs dimension 3). The most noticeable aspect is the similarity between the layouts of the 7 cities in these two spaces. In addition, panel (c) compares the ranks of pairwise distances among the 7 cities in the geographical space with those in the latent space. The correlation coefficient between the two sets of ranks is as high as 0.96. The above findings suggest that geographical layout is an important factor in explaining the contemporaneous dependence between the 7 cities' temperature movements. This conclusion is obviously not unexpected since neighboring cities are likely to share similar meteorological and geological conditions, which will impact the temperature within a region. Here we emphasize that {\em no} geographical information is provided to our model. The latent positions, as given by the rows of $\hat{U}$ as in \eqref{Uest}, are discovered purely by the reduced-rank covariance estimation in the VAR modeling of the temperature data. 

For model diagnostics, panel (d) of Figure \ref{cityTemperature_rr_combine} displays the ACF and CCF among the 3 dimensions of the estimated latent variable $\hat{\delta}_{t}$ as computed in \eqref{reduceRankDecomEstDelta}. We can see that, with few exceptions, neither the auto-correlation nor the cross-correlation is significant, which is consistent with the model assumptions. 

\newpage
~\\\noindent{\bf Acknowledgments}. We would like to thank Professor Qiwei Yao for providing the temperature data. The research of Richard A. Davis is supported in part by NSF grant DMS-1107031. The research of Tian Zheng is supported in part by NSF grant SES-1023176 and a 2010 Google faculty research award.

\section{Appendix}\label{section_appendix_rr}
\subsection{Proof of Proposition \ref{reduceRankDecomEstProp} in Section \ref{section_RR_independent_rr}}\label{section_appendix_proof_rr}
{\bf Proof of Proposition \ref{reduceRankDecomEstProp}}. Notice that the $K$ eigenvalues of $\Sigma_{Z}=U\Lambda U^{'} + \sigma^2I_{K}$ are $\lambda_{1}+\sigma^2,\ldots,\lambda_{d}+\sigma^2,\sigma^2,\ldots,\sigma^2$, so the $-\frac{2}{T}$log-likelihood \eqref{logLikelihood} becomes
\begin{align}
-\frac{2}{T}\log L(U,\Lambda, \sigma^2) &=  \log|\Sigma_{Z}| + \mathrm{tr}(\Sigma_{Z}^{-1}S) \nonumber \\
                                                           &= (K - d)\log(\sigma^2) + \displaystyle\sum_{i=1}^{d}\log(\lambda_i + \sigma^2)+\mathrm{tr}(\Sigma_{Z}^{-1}S). \label{negLogLikelihood}
\end{align}
From standard matrix results, see e.g., \citet{Schott2004}, \eqref{reduceRankCovMatr} gives
\begin{eqnarray}
\Sigma_{Z}^{-1} &=& (U\Lambda U^{'} + \sigma^2I_K)^{-1} \nonumber\\
            &=& (\sigma^2I_K)^{-1} - (\sigma^2I_K)^{-1}U[\Lambda^{-1} + U^{'}(\sigma^2I_K)^{-1}U]^{-1}U^{'}(\sigma^2I_K)^{-1} \nonumber \\
            &=& \frac{1}{\sigma^2}I_K - \frac{1}{(\sigma^2)^2}U(\mathrm{diag}\{\frac{1}{\lambda_1}, \frac{1}{\lambda_2},\ldots,\frac{1}{\lambda_d}\} + \mathrm{diag}\{\frac{1}{\sigma^2}, \frac{1}{\sigma^2},\ldots,\frac{1}{\sigma^2}\})^{-1}U^{'}\nonumber\\
            &=& \frac{1}{\sigma^2}I_K + \frac{1}{\sigma^2}U\mathrm{diag}\{-\frac{\lambda_1}{\lambda_1 + \sigma^2},-\frac{\lambda_2}{\lambda_2 + \sigma^2},\ldots,-\frac{\lambda_d}{\lambda_d + \sigma^2}\}U^{'}\nonumber\\
            &=& \frac{1}{\sigma^2}(I_K + U\tilde{\Lambda}U^{'}),
\label{negLogLikelihood_1}
\end{eqnarray}
where $\tilde{\Lambda} \defeq \mathrm{diag}\{-\frac{\lambda_1}{\lambda_1 + \sigma^2},\ldots,-\frac{\lambda_d}{\lambda_d + \sigma^2}\}$. We point out that it is the assumption of the isotropic error covariance matrix $\mathrm{var}(\varepsilon_{t})=\sigma^2 I_{K}$ that makes it possible to explicitly calculate $\Sigma_{Z}^{-1}$ as in \eqref{negLogLikelihood_1} and eventually leads to the analytical form of the maximum likelihood estimator. Plugging \eqref{negLogLikelihood_1} into \eqref{negLogLikelihood}, we have
\begin{align}
-\frac{2}{T}\log L(U,\Lambda, \sigma^2) &= (K - d)\log(\sigma^2) + \displaystyle\sum_{i=1}^{d}\log(\lambda_i + \sigma^2) + \frac{1}{\sigma^2}\mathrm{tr}[(I_K + U\tilde{\Lambda}U^{'})S] \nonumber \\
            &= (K - d)\log(\sigma^2) + \displaystyle\sum_{i=1}^{d}\log(\lambda_i + \sigma^2) + \frac{1}{\sigma^2}\mathrm{tr}(S) + \frac{1}{\sigma^2}\mathrm{tr}(U\tilde{\Lambda}U^{'}S)\nonumber \\
 &= (K - d)\log(\sigma^2) +  \displaystyle\sum_{i=1}^{d}\log(\lambda_i + \sigma^2) + \frac{1}{\sigma^2}\displaystyle\sum_{i=1}^{K}c_{i} + \frac{1}{\sigma^2}\mathrm{tr}(U^{'}SU\tilde{\Lambda}) \label{negLogLikelihood_2}.
\end{align}
Let $\hat{U}$ denote the $K\times d$ matrix whose columns consist of the $d$ eigenvectors that correspond to the $d$ largest eigenvalues of $S$ as in \eqref{Uest}. Since the diagonal entries of $\tilde{\Lambda}$ are negative and in increasing order, i.e.,$-\frac{\lambda_1}{\lambda_1 + \sigma^2} < \ldots < -\frac{\lambda_d}{\lambda_d + \sigma^2} < 0$,
standard matrix results, e.g., see \citet{Horn2013},  show that $\mathrm{tr}(U^{'}SU\tilde{\Lambda})$ in \eqref{negLogLikelihood_2} is minimized by $\hat{U}$. In addition, as long as the relationship $-\frac{\lambda_1}{\lambda_1 + \sigma^2} < \ldots < -\frac{\lambda_d}{\lambda_d + \sigma^2} < 0$ holds, $\hat{U}$ is the minimizer regardless of the particular values of $\lambda_{1},\ldots,\lambda_{d}$ and $\sigma^2$.
If the $d$ largest eigenvalues $c_{1},\ldots,c_{d}$ of $S$ are distinct, the minimizer $\hat{U}$ is
unique up to column-wise reflections. Additionally, $\hat{U}$ is unique if the signs of entries in one row of $\hat{U}$ are anchored {\em a priori}. 

Now we have $\hat{U}^{'}S\hat{U} = \mathrm{diag}\{c_1,\ldots,c_d\}$, so plugging $\hat{U}$ into \eqref{negLogLikelihood_2} gives
\begin{align}
-\frac{2}{T}\log L(\hat{U},\Lambda, \sigma^2) &= (K-d)\log(\sigma^2) + \displaystyle\sum_{i=1}^{d} \log(\lambda_i + \sigma^2) 
+  \frac{1}{\sigma^2}\displaystyle\sum_{i=1}^{K}c_i + \frac{1}{\sigma^2}\mathrm{tr}(\mathrm{diag}\{c_1,\ldots,c_d\}\tilde{\Lambda}) \nonumber \\
&= (K-d)\log(\sigma^2) +\displaystyle\sum_{i=1}^{d} \log(\lambda_i + \sigma^2) + \frac{1}{\sigma^2}\displaystyle\sum_{i=1}^{K}c_i - \frac{1}{\sigma^2}\displaystyle\sum_{i=1}^{d}\frac{\lambda_ic_i}{\lambda_i + \sigma^2} \nonumber \\
                                          &= (K-d)\log(\sigma^2) + \displaystyle\sum_{i=1}^{d} \log(\lambda_i + \sigma^2) + \frac{1}{\sigma^2}\displaystyle\sum_{i=d+1}^{K}c_i + \displaystyle\sum_{i=1}^{d}\frac{c_i}{\lambda_i + \sigma^2.} \label{negLogLikelihood_3}
\end{align}
Minimizing the right-hand size of \eqref{negLogLikelihood_3} with respect to $\lambda_{1},\ldots,\lambda_{d}$ and $\sigma^2$, we have
\begin{eqnarray}
\hat{\sigma}^2 &=& \frac{1}{K-d}\displaystyle\sum_{i=d+1}^{K}c_i; \nonumber\\
\hat{\lambda}_{i} &=& c_i - \hat{\sigma}^2,\mbox{ for } i=1,\ldots,d. \nonumber
\end{eqnarray}
which completes the proof.

\subsection{Approximation of MSE matrices of VAR forecasting}\label{section_appendix_mse_rr}
We give results on approximating the mean squared error (MSE) matrix for one-step forecast of a VAR model. Let $\{Y_{t}\}$ be the VAR($p$) process in \eqref{VAR_equation} with $\mu={\bf 0}$. Then the optimal one-step forecast of $Y_{t}$ based on $Y_{t},\ldots,Y_{1}$ with {\em estimated} AR coefficients $\hat{A}_{1},\ldots,\hat{A}_{p}$ is given by
\begin{eqnarray}
\hat{Y}_{t}(1) = \displaystyle\sum_{k=1}^{p}\hat{A}_{k}Y_{t+1-k}, ~\mbox{ for } t>p. \nonumber
\end{eqnarray}
It can be shown, see e.g., \citet{Helmut1991}, that the MSE matrix of the 1-step forecast $\hat{Y}_{t}(1)$, which is defined as
\begin{eqnarray}
\mathrm{fMSE}(1) \defeq \mathbb{E}[Y_{t+1}-\hat{Y}_{t}(1)][Y_{t+1}-\hat{Y}_{t}(1)]^{'}, \nonumber
\end{eqnarray}
can be approximated by
\begin{eqnarray}
\tilde{\mathrm{fMSE}}(1)\defeq \Sigma_{Z} + \Omega(1), \label{oneStepMSEapprox} 
\end{eqnarray}
where
\begin{align}
\Omega(1) &\defeq \frac{1}{T}\displaystyle\sum_{t=1}^{T}\{(L_{t}^{'}\Gamma_{Y}^{-1}L_{t})\otimes\Sigma_{Z}\}, \label{oneStepMSE_part3} \\
L_{t}      &\defeq (Y_{t}, Y_{t-1},\ldots, Y_{t-p+1})^{'},\mbox{ for } t=1,\ldots,T, \label{oneStepMSE_part1}\\
\Gamma_{Y} &\defeq \mathrm{cov}(L_{t}) = \mathrm{cov}(Y_{t}, Y_{t-1},\ldots, Y_{t-p+1})^{'}. \label{oneStepMSE_part2}
\end{align}
We can see that the approximate one-step forecast MSE matrix $\tilde{\mathrm{fMSE}}(1)$ \eqref{oneStepMSEapprox} has two parts: the first part $\Sigma_{Z}$ comes from the uncertainty inherent in the VAR model while the second part $\Omega(1)$ given in \eqref{oneStepMSE_part3} accounts for the variability in the parameter estimates. We estimate the approximate one-step forecast MSE matrix $\tilde{\mathrm{fMSE}}(1)$ by plugging the parameter estimates $\hat{A}_{1},\ldots,\hat{A}_{p}$ and $\hat{\Sigma}_{Z}$ into \eqref{oneStepMSEapprox}. 
For such estimation, we need to represent the $Kp\times Kp$ covariance matrix $\Gamma_{Y}=\mathrm{cov}(Y_{t}, Y_{t-1},\ldots, Y_{t-p+1})^{'}$ \eqref{oneStepMSE_part2} in terms of $A_{1},\ldots,A_{p}$ and $\Sigma_{Z}$. We derive this representation as follows. From \eqref{VAR_equation} with $\mu={\bf 0}$, we can see that the $Kp$-dimensional process $\{L_{t}\}$ \eqref{oneStepMSE_part1} satisfies the following VAR(1) recursion $L_{t}=\Psi L_{t-1}+V_{t}$, i.e.,
\begin{align}\label{VARexpand}
\left(\begin{array}{c}
Y_{t}  \\
Y_{t-1}  \\
Y_{t-2} \\
\vdots  \\
Y_{t-p+1}  
\end{array}\right)        
&= 
\left(\begin{array}{ccccc}
A_1 & A_2 & \cdots & \cdots & A_p \\
I_K & 0 & \cdots & \cdots & 0 \\
0 & I_K & \cdots & \cdots & 0 \\
\vdots & \vdots & \vdots & \vdots & \vdots \\
0 & 0 & \cdots & I_K & 0\\
\end{array}\right) 
\left(\begin{array}{c}
Y_{t-1}  \\
Y_{t-2}  \\
Y_{t-3} \\
\vdots  \\
Y_{t-p}  
\end{array}\right)  
+
\left(\begin{array}{c}
Z_{t}  \\
0  \\
0  \\
\vdots  \\
0
\end{array}\right), 
\end{align}
where the $Kp\times Kp$ AR coefficient matrix $\Psi$ in \eqref{VARexpand} is referred to as the {\em companion matrix} of the VAR($p$) model \eqref{VAR_equation}, e.g., see \citet{Helmut1991}. The covariance matrix $\Sigma_{V}$ of the $Kp$-dimensional noise $V_{t}$ in \eqref{VARexpand} is a $Kp\times Kp$ matrix of zeros except its upper-left $K\times K$ sub-matrix being equal to $\Sigma_{Z}$. From \eqref{oneStepMSE_part2} and the VAR(1) recursion $L_{t}=\Psi L_{t-1}+V_{t}$, we can see that
\begin{eqnarray}
\Gamma_{Y} = \mathrm{cov}(L_{t}) = \mathrm{cov}(\Psi L_{t-1} + V_{t}) = \Psi\Gamma_{Y}\Psi^{'} + \Sigma_{V}, \label{Gamma_1}
\end{eqnarray}
and \eqref{Gamma_1} leads to
\begin{align}
\mathrm{vec}(\Gamma_{Y}) &= \mathrm{vec}(\Psi\Gamma_{Y}\Psi^{'} + \Sigma_{V}) \nonumber\\  
                                   &= \mathrm{vec}(\Psi\Gamma_{Y}\Psi^{'}) + \mathrm{vec}(\Sigma_{V}) \nonumber\\
                                   &= (\Psi\otimes\Psi)\mathrm{vec}(\Gamma_{Y}) + \mathrm{vec}(\Sigma_{V}). \label{Gamma_2}
\end{align} 
From \eqref{Gamma_2}, it follows that
\begin{eqnarray}
\mathrm{vec}(\Gamma_{Y}) = (I_{K^2p^2} - \Psi\otimes\Psi)^{-1}\mathrm{vec}(\Sigma_{V}). \label{Gamma_3}
\end{eqnarray}
Replacing $A_{1},\ldots,A_{p}$ and $\Sigma_{Z}$ with their estimates in \eqref{Gamma_3}, we can obtain estimates for the $Kp\times Kp$ covariance matrix $\Gamma_{Y}$ \eqref{oneStepMSE_part2} and thereby estimates for the approximate one-step forecast MSE matrix $\tilde{\mathrm{fMSE}}(1)$ \eqref{oneStepMSEapprox}.

\bibliographystyle{asa}
\bibliography{paperForRRCovMatrix_JEcon}
\end{document}